\documentclass{article}
\usepackage{afterpage}

\addtocontents{toc}{\protect\setcounter{tocdepth}{-1}}

\usepackage{microtype}
\usepackage{graphicx}
\usepackage{subcaption}
\usepackage{booktabs} 

\usepackage{hyperref}
\usepackage{makecell}
\usepackage{placeins}

\usepackage[ruled, vlined, linesnumbered]{algorithm2e}

\usepackage[accepted]{icml2026}

\usepackage{amsmath}
\usepackage{amssymb}
\usepackage{mathtools}
\usepackage{amsthm}
\usepackage{enumitem}
\usepackage{booktabs}
\usepackage{pifont}
\usepackage{xspace}
\usepackage{multirow}

\usepackage{xcolor}
\usepackage{pifont}
\usepackage[table]{xcolor}
\usepackage{tcolorbox}
\usepackage{tabularx}

\definecolor{checkgreen}{RGB}{34,139,34} 
\definecolor{crossred}{RGB}{178,34,34}   

\newcommand{\cmark}{\textcolor{checkgreen}{\ding{51}}}
\newcommand{\xmark}{\textcolor{crossred}{\ding{55}}}

\usepackage[capitalize,noabbrev]{cleveref}

\theoremstyle{plain}

\theoremstyle{definition}

\theoremstyle{remark}

\definecolor{darkgreen}{RGB}{0,100,0}

\newcommand{\model}{Clean-PR\xspace}
\usepackage[textsize=tiny]{todonotes}

\icmlsetsymbol{equal}{*}

\renewcommand{\icmlEqualContribution}{\textsuperscript{*}Work done during internship at Microsoft Research. 
Qinglin Zhu is supervised by Lin Gui and Yulan He.}

\begin{document}

\twocolumn[
  \icmltitle{ Pull Requests as a Training Signal for Repo-Level Code Editing}




\begin{icmlauthorlist}
  \icmlauthor{Qinglin Zhu}{equal,kcl}
  \icmlauthor{Tianyu Chen}{comp}
  \icmlauthor{Shuai Lu}{comp}
  \icmlauthor{Lei Ji}{comp}
  \icmlauthor{Runcong Zhao}{kcl}
  \icmlauthor{Murong Ma}{nus}
  \icmlauthor{Xiangxiang Dai}{cuhk}
  \icmlauthor{Yulan He}{kcl,turing}
  \icmlauthor{Lin Gui}{kcl}
  \icmlauthor{Peng Cheng}{comp}
  \icmlauthor{Yeyun Gong}{comp}
\end{icmlauthorlist}

\icmlaffiliation{kcl}{King's College London, UK}
\icmlaffiliation{cuhk}{Chinese University of Hong Kong, HK}
\icmlaffiliation{nus}{National University of Singapore, SG}
\icmlaffiliation{comp}{Microsoft Research Asia, CN}
\icmlaffiliation{turing}{The Alan Turing Institute, UK}
\icmlcorrespondingauthor{Tianyu Chen}{chentianyu@microsoft.com}

  \icmlkeywords{Machine Learning, ICML}

  \vskip 0.3in
]



\printAffiliationsAndNotice{%
\icmlEqualContribution\quad}

\begin{abstract}
Repository-level code editing requires models to understand complex dependencies and execute precise multi-file modifications across a large codebase. While recent gains on SWE-bench rely heavily on complex agent scaffolding, it remains unclear how much of this capability can be internalised via high-quality training signals. Thus, we propose \textbf{Clean Pull Request (\model)}, a mid-training paradigm that leverages real-world GitHub pull requests as a training signal for repository-level editing. 
We introduce a scalable pipeline that converts noisy PR diffs into Search/Replace edit blocks by reconstruction and validation, yielding the largest publicly available corpus of \textbf{2M PRs} across \textbf{12 languages}.
Leveraging this signal, we implement a mid-training stage, followed by an Agentless-aligned SFT with error-driven augmentation. 
On SWE-bench, our model significantly outperforms the instruct model baseline, achieving absolute gains of 13.6\% on Lite and 12.3\% on Verified. 
Our results show that repo-level capabilities can be internalised into model weights under a simplified Agentless protocol, reducing reliance on heavy inference scaffolding.

\end{abstract}

\section{Introduction}

Repository-level software engineering (SWE) has emerged as a crucial testbed for code-capable large language models (LLMs), driven by executable benchmarks such as SWE-bench \citep{jimenez2024swebench}. 
State-of-the-art SWE systems typically rely on composite architectures, combining agentic tool use \citep{yang2024sweagent,wang2024openhands}, structured localisation \citep{xia2024agentless,jiang2025cosil}, and extensive test-time scaling \citep{antoniades2024swesearch}. 
While effective, this complexity makes it difficult to attribute performance gains to any single factor.
This motivates a fundamental question: \emph{how much repository-level editing capability can be encoded directly into model weights?}

Answering this question requires abundant, high-quality training data aligned with repository-level editing. 
However, as summarised in Table~\ref{tab:dataset_comparison}, current data landscapes present a dichotomy. 
On one hand, SWE-bench-style datasets provide high-fidelity, executable verification but are expensive to curate and limited in scale \citep{pan2025swegym,jain2025r2egym}. 
On the other hand, massive pretraining corpora like The Stack \citep{kocetkov2022stack} and CodeReview~\cite{li2022automatingcodereviewactivities} offer scale but lack instruction on \emph{how} to modify codebases to resolve issues. 
There remains a clear gap in datasets that combine the scale of natural corpora with the structured, multi-file editing signals required for SWE tasks

Open-source Pull Requests (PRs) offer a promising middle ground, naturally coupling human intent expressed in natural language (descriptions, discussions) with accepted code changes, thereby providing a rich training signal for both \emph{where to edit} and \emph{how to edit}. 
Yet, raw PR traces (visualised in Table~\ref{tab:raw_data_instances}) are a noisy proxy for high-quality training data. 
As shown in Table~\ref{tab:filter_stats}, a naïve ingestion of GitHub PRs results in substantial noise: a significant majority are discarded due to being bot-generated, lacking core source code changes, or simply remaining unmerged.
This motivates a rigorous conversion pipeline that turns PRs into model-ready and scalable training data for repository-level editing.

\begin{table*}[!htbp]
\caption{Comparison of repo-level SWE datasets and natural code-edit corpora.}
\vspace{0.1in}
\centering
\resizebox{0.88\linewidth}{!}{
\begin{tabular}{lccccrr}
\toprule[1pt]
Dataset 
& Real task 
& Multi-file
& Diff format 
& \#Lang 
& \#Repo
& \#Instances \\
\midrule

\multicolumn{7}{l}{\textit{SWE-bench-style datasets: }} \\
Multi-SWE-bench~\cite{zan2025multiswebench} & \cmark & \cmark & Diff & 7 & 39 & 1,632 \\
SWE-PolyBench~\cite{rashid2025swepolybench} & \cmark & \cmark & Diff & 4 & 21 & 2,110 \\
SWE-bench-Live~\cite{zhang2025swebenchlive} & \cmark & \cmark & Diff & 1 & 164  & 1,565 \\
SWE-Gym~\cite{pan2025swegym}                & \cmark & \cmark & Diff & 1 & 11 & 2,438 \\
R2E-Gym~\cite{jain2025r2egym}               & \cmark & \cmark & Diff & 1 & 10 & 8,135 \\
SWE-rebench~\cite{badertdinov2025swerbench} & \cmark & \cmark & Diff & 1 & 3468 & 21,324 \\
SWE-smith~\cite{yang2025swesmith}           & \xmark & \cmark & Diff & 1 & 128 & 50,000 \\
SWE-Synth~\cite{pham2025swesynthsynthesizingverifiablebugfix} & \xmark & \cmark & Diff & 1 & 7 & 9,459 \\
\midrule
\multicolumn{7}{l}{\textit{Natural code-change corpora:}} \\
The Stack~\cite{kocetkov2022stack} & \cmark & \xmark & None  & 30 & -- & 317M \\
commitpackft~\cite{muennighoff2023octopack} & \cmark & \xmark & Before/After & 277 & -- & 742,273 \\
commitbench~\cite{schall2024commitbenchbenchmarkcommitmessage} & \cmark & \xmark & Diff & 6 & 72,000 & 1,165,213 \\
CodeReview~\cite{li2022automatingcodereviewactivities} & \cmark & \xmark & Diff & 9 & 1,161 & 534,000 \\
\midrule
\textbf{\model-full (ours)} 
& \textbf{\cmark} 
& \textbf{\cmark}
& Search/Replace 
& \textbf{12} 
& \textbf{52,338}
& \textbf{3,050,939} \\
\textbf{\model-train (ours)} 
& \textbf{\cmark} 
& \textbf{\cmark}
& Search/Replace 
& \textbf{12} 
& \textbf{45,267}
& \textbf{2,015,708} \\
\bottomrule[1pt]
\end{tabular}
}
\vspace{0.05in}
\label{tab:dataset_comparison}
\end{table*}

\begin{figure*}
    \centering
    \includegraphics[width=1 \linewidth]{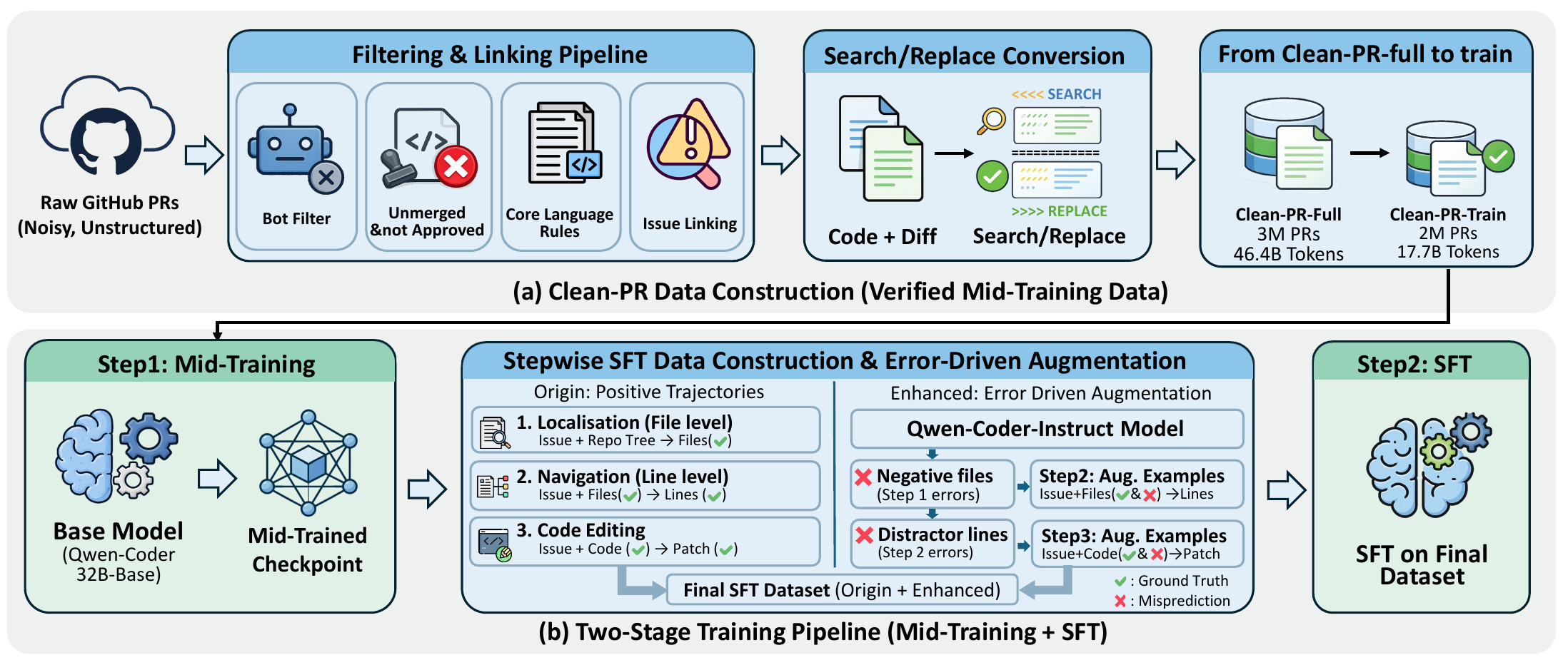}
    \caption{\textbf{Overview of the \model Framework.} \textbf{(a) Data Construction:} Raw GitHub PRs undergo a rigorous filtering pipeline (bot detection, core language enforcement) and intent augmentation via linked Issues. The valid diffs are then converted into minimal unique \textbf{Search/Replace} blocks, verified through round-trip patch application to ensure correctness. \textbf{(b) Two-Stage Training Pipeline:} The base model first undergoes \textbf{Mid-Training} on the verifiable \model corpus to encode repository-level editing priors. This is followed by an \textbf{Agentless-Aligned Stepwise SFT}, where the model is fine-tuned on decomposed tasks (Localisation $\rightarrow$ Navigation $\rightarrow$ Editing) with \textbf{Error-Driven Augmentation} to robustly handle distracting repository contexts.}
    \label{fig:overview}
    \vskip -0.2in
\end{figure*}

In this work, we propose \textbf{\model}, a scalable mid-training paradigm that transforms noisy PRs into a rigorous training signal for repository-level editing (Figure~\ref{fig:overview}). 
To ensure high-quality supervision at scale, we implement a data construction pipeline consisting of three main steps.
\textbf{First, Noise Filtering and Issue Linking:} We design a high-precision pipeline to filter low-signal PRs and, given that PR and issue content are stored separately, we detect referenced identifiers to augment examples with faithful issue context.
\textbf{Second, Search/Replace Conversion:} We reconstruct repository-consistent \emph{before/after} states and \emph{verify} that the derived Search/Replace blocks deterministically reproduce the post-PR code exactly.
In particular, we utilise Search/Replace edit blocks rather than standard diffs to align with widely adopted code-editing pipelines~\citep{xia2024agentless, wang2024openhands},
avoiding the fragility of diffs where valid application hinges on precise line-number prediction (Table~\ref{tab:diff_vs_sr}).
\textbf{Third, Downstream Sampling:} This process yields a large-scale corpus of approximately \textbf{3M PRs} spanning \textbf{12} programming languages (46.4B tokens), from which we select a \textbf{17.7B} token subset for mid-training by applying modification scope constraints and balanced repository sampling.
To the best of our knowledge, this is the \textbf{largest publicly released PR-derived dataset} explicitly constructed and validated for repo-level mid-training.

However, mid-training alone does not fully address the challenge of robust localisation and navigation within massive repositories.
To bridge this gap, we implement two targeted strategies.
First, we align the model with the inference protocol via \textbf{Agentless-aligned SFT}, training it to explicitly decompose the problem into (i) file selection, (ii) line-level navigation, and (iii) executable Search/Replace patch generation.
Second, to prevent \emph{over-editing}~\cite{zeng2025hyperedit} caused by the model's failure to reject irrelevant files within noisy retrieval results, we introduce an \textbf{Error-Driven Augmentation} strategy.
By injecting distractor files and regions mined from incorrect predictions, we teach the model to reject irrelevant context.
This recipe yields consistent improvements on SWE-bench, enhancing both intermediate localisation metrics and end-to-end repair success.

Our contributions can be summarised as follows:
\begin{itemize}[noitemsep]
    \item \textbf{\model Framework:} We propose a data-centric mid-training paradigm transforming noisy GitHub PRs into rigorous training signal. By implementing strict noise filtering and verifying targets through patch application, we bridge the gap between open-source noise and precise repository editing.
    \item \textbf{Largest Verified PR Corpus:} We plan to release the  verifiable PR dataset to date (2M PRs). Unlike standard diffs, it utilises a robust Search/Replace format augmented with issue contexts, explicitly constructed to support high-fidelity repository editing tasks.
    \item \textbf{Agentless-Aligned Training:} We propose  an SFT strategy with error-driven augmentation to align the model with a simplified Agentless workflow, enabling it to discriminate against distracting context and navigate code without relying on complex agentic loops.
    \item \textbf{Comprehensive Evaluation:} Applying our method to the Qwen2.5-Coder-32B base, we achieve absolute gains of 13.6\% on SWE-bench Lite and 12.3\% on Verified over baselines. Additional evaluations on 7B models, multilingual repair, and OpenHands show that robust engineering capabilities can be encoded into model weights and transfer across inference paradigms.
\end{itemize}

\begin{table}[t]
\caption{Statistics of noise categories applied to the raw dataset. Categories are not mutually exclusive: a single PR may match multiple patterns (e.g., a bot-created PR that is also unmerged).
}
\label{tab:filter_stats}
\centering
\begin{small}
\resizebox{0.9 \linewidth}{!}{%
\begin{tabular}{lrr}
\toprule
\textbf{Noise Category} & \textbf{Count} & \textbf{Ratio (\%)} \\
\midrule

Non-Core Source Changes & 6,245,784 & 38.06 \\
Suspected Robot Activity & 4,112,501 & 25.10 \\
Unmerged / Not Approved & 4,016,574 & 24.50 \\
Empty Base File or Diff & 2,241,453 & 13.66 \\
Patch Validation Failure & 675,825 & 4.10 \\
\midrule

\rowcolor{green!15} \textbf{Clean (w/o Noise)} & \textbf{3,050,939} & \textbf{18.59} \\
\textbf{Total (Raw PR)} & \textbf{16,408,886} & \textbf{100.00} \\
\bottomrule
\end{tabular}%
}
\end{small}
\end{table}

\section{Data Construction}
\label{sec:data}

Our objective is to encode repository-level editing capabilities directly into model weights, minimising reliance on heavy inference-time scaffolding. 
While open-source pull requests (PRs) offer a massive source of developer intent coupled with code changes, raw PR traces are often dominated by noise.
To bridge the gap between noisy wild data and the rigorous requirements of repository editing, we construct our data in two stages:
(1) \textbf{\model}, a verified mid-training corpus derived from millions of filtered and reconstructed PRs, and 
(2) an \textbf{Agentless-Aligned SFT} dataset designed to bridge the gap between pure editing and the multi-stage localisation-then-editing workflow required by benchmarks like SWE-bench.

\subsection{\model: Verified Mid-Training Data}
\label{subsec:mid_training}

We implement a rigorous pipeline to transform raw GitHub activity into a verifiable training signal. 
We illustrate a concrete example of a constructed data instance in Figure~\ref{fig:appendix_pipeline_visual} and provide detailed processing specifications in Appendix~\ref{app:data_processing}.

\paragraph{Data Collection.}
To construct a comprehensive corpus for repository-level editing, we undertook a crawl of publicly available GitHub pull requests. At the time of this work, our collection comprises approximately \textbf{8.6 TB} of raw data, spanning 274k repositories and 16.4 million PRs.

\paragraph{Data Filtering.}
To assess the suitability of raw GitHub data for training, we conducted a preliminary study on the noise distribution within our initial collection.
Specifically, we defined a noise taxonomy by synthesizing heuristics from \citep{lozhkov2024starcoder2stackv2, kocetkov2022stack} with manual inspection.
As detailed in Table~\ref{tab:filter_stats}, the raw stream is heavily polluted: 38.06\% of PRs lack core source changes, 25.10\% stem from bot activity, and 24.50\% remain unmerged.
Motivated by these findings, we apply a rigorous filtering protocol to aggressively prune these non-learning signals, retaining only \textbf{18.59\%} of the original data as high-density training signals.
Crucially, to prevent data contamination, we explicitly exclude all repositories present in the SWE-bench evaluation sets from our final corpus.

\begin{itemize}[noitemsep]
    \item \textbf{PR Validity:} Raw GitHub data contains substantial noise. We discard PRs that are unmerged, closed without merging, or created solely by automated accounts (bots). We further exclude PRs that exclusively change documentation, which are unsuitable for learning focused semantic edits.
    \item \textbf{Language Alignment:} Repository-level training must emphasise semantic code editing rather than configuration churn. We enforce a \textit{Core Extension Rule}, retaining a PR only if it modifies at least one core source file corresponding to our 12 target languages, with specific definitions provided in Table~\ref{tab:language_rules}. PRs dominated by configuration or auto-generated files are removed to ensure the model learns meaningful logic changes.
\end{itemize}

\paragraph{Search/Replace format Reconstruction.}
Standard unified diffs are brittle for LLM generation due to their reliance on fragile line numbers. We instead utilise the \textbf{Search/Replace} format \citep{xia2024agentless}, which locates edits via unique context matching. 
To ensure these targets are grounded in a valid repository state, we perform a rigorous round-trip verification. 
The detailed algorithmic procedure is outlined in Algorithm~\ref{alg:sr_gen_detailed}.
First, we reconstruct the exact ``after'' state of the repository by applying the raw PR patch to the ``before'' snapshot. 
We then algorithmically derive minimal edit spans and select unique anchor contexts to form Search/Replace blocks. Finally, we \emph{verify} these blocks by applying them back to the ``before'' state; any example where the re-application does not bit-wise match the ground-truth ``after'' state is discarded. This guarantees that every training example corresponds to a \textbf{minimal unique search block}, defined as the shortest contiguous span of context needed to uniquely identify the edit location within a file, thereby filtering out noisy diffs caused by formatting drift.

\begin{table}[t]
\caption{\model mid-training corpus statistics (stage-wise).}
\label{tab:midtrain_stats}
\centering
\resizebox{\linewidth}{!}{
\setlength{\tabcolsep}{5pt}
\begin{tabular}{lrr}
\toprule
\textbf{PR Cleaning Stage} & \textbf{\#PRs} & \textbf{Tokens} \\
\midrule
Raw PR snapshot (with code) & 16,408,886 & -- \\
Filtering + S/R format (\textbf{\model-full}) & 3,050,939 & 46.4B \\
Core-file constraint ($\le 5$ core files) & 2,751,937 & 29.7B \\
After format/length constraints & 2,585,778 & 22.0B \\
\textbf{\model-train (repo-sampling)} & \textbf{2,015,708} & \textbf{17.7B} \\
\bottomrule
\end{tabular}
}
\end{table}

\begin{table}[t]
\caption{Detailed statistical comparison between the full verified corpus and the final training set. }
\label{tab:data_comparison}
\centering
\resizebox{ \linewidth}{!}{
\begin{tabular}{lrr}
\toprule
\textbf{Metric} & \textbf{\model-full} & \textbf{\model-train} \\
\midrule
Total Instances & 3,050,939 & 2,015,708 \\
Total Repositories & 52,338 & 45,267 \\
Total Tokens & 46.4B & 17.7B \\
\midrule
Avg. Description Len (words) & 50.0 & 59.5 \\
Avg. Modified Files & 3.0 & 1.7 \\
Avg. Code Lines & 1,562.3 & 1,077.8 \\
Avg. Search/Replace Blocks & 9.0 & 4.3 \\
Avg. Search/Replace Lines & 130.3 & 58.7 \\
Avg. Comments & 2.2 & 2.1 \\
\bottomrule
\end{tabular}
}
\end{table}

\paragraph{Issue-Augmented Intent.}

Raw PR descriptions frequently lack self-contained context, as developers customarily reference an external Issue identifier (e.g., ``Fixes \#123'') rather than restating the detailed bug report or feature requirement. 
To recover this missing problem definition, we implement an issue augmentation pipeline: we retrieve and concatenate the titles and descriptions of all linked issues into the PR context.
By supplementing brief developer summaries with original user reports, we align the training signal more closely with real-world software engineering workflows, where the initial prompt is typically a detailed bug report rather than a known solution.

\paragraph{From \model-full to \model-train.}
As detailed in Table~\ref{tab:midtrain_stats}, our initial filtering yields the \textbf{\model-full} corpus, comprising 3.05 million verified instances (46.4B tokens). 
To construct the final \textbf{\model-train} dataset, we apply three targeted refinements. 
First, to focus the model on self-contained, learnable units of work, we implement \textbf{complexity control} by restricting the dataset to PRs modifying at most five core files; as shown in Table~\ref{tab:data_comparison}, this reduces the average modified files from 3.0 to 1.7. 
Second, we apply \textbf{context windowing} to files exceeding 100k tokens, preserving the editing signal by centring the input window around verified Search/Replace blocks. 
Finally, to mitigate distribution skew from dominant projects, we enforce \textbf{repository-level sampling}: for any repository contributing more than 2,000 PRs, we randomly sample exactly 2,000 instances, while retaining all PRs from smaller repositories, resulting in a balanced corpus of \textbf{2 million} instances.
The detailed language distribution is provided in Appendix~\ref{app:Language_Distribution}, with Python PRs accounting for 19.3\% of the training set.

\subsection{Agentless-Aligned Stepwise SFT}
\label{subsec:sft}

Mid-training equips the model with the fundamental capability to edit code given a context. However, resolving repository-scale issues necessitates a structured \textbf{decompose-and-solve} workflow: locating files, identifying regions, and then editing a paradigm established as robust by recent systems~\citep{xia2024agentless, yang2024sweagent, jiang2025cosil}.
To bridge the gap between our raw editing capability and this structured requirement, we construct a high-quality SFT dataset aligned with a Simplified Agentless workflow. We favour this generative approach over complex Agentic frameworks as it leverages the strong reasoning priors of our base model without the overhead of multi-turn dialogue or external tool use.
We achieve this by leveraging verified trajectories from SWE-rebench~\cite{badertdinov2025swerbench} and SWE-Gym~\citep{pan2025swegym}, effectively ``baking" fine-grained navigation intelligence directly into the model weights to reduce reliance on external scaffolding.

\paragraph{Task Decomposition and Filtering.}
We derive training samples by decomposing each ground-truth repair into three distinct supervised tasks, rigorously filtering for tractability:

\begin{itemize}[noitemsep]
    \item \textbf{File localisation} (Issue + Repo Tree $\rightarrow$ Filepath): 
    Navigating real-world codebases is a needle-in-a-haystack challenge; for instance, repositories in SWE-bench contain an average of 3,010 files, yet a typical issue requires modifying only 1.7 files~\citep{jimenez2024swebench}. Therefore, effectively narrowing the search space is a prerequisite for tractable editing. In this step, the model learns to identify relevant source files solely from the issue description and repository structure. To focus the training signal on functional repairs, we filter the target labels to exclude non-code artefacts (e.g., \texttt{.md}, \texttt{.txt}), ensuring the model learns to prioritise semantic code changes over auxiliary file updates.
    \item \textbf{Fine-grained Navigation} (Issue + File Content $\rightarrow$ Relevant Context): 
    Since source files often span thousands of lines, operating on full-file context is inefficient and prone to distraction. We utilise Abstract Syntax Tree (AST) parsing to map ground-truth edits to their enclosing function or class definitions. This trains the model to identify the precise logical scope of the bug, rather than arbitrary line numbers, ensuring robustness against minor formatting shifts.
    \item \textbf{Patch Generation} (Localised Context $\rightarrow$ Search /Replace Patch): 
    Given the identified code region, the model generates the final fix. We enforce that the target output forms a \textbf{minimal unique search block}, which prevents ambiguous application errors while minimising token usage compared to verbose context blocks.

\end{itemize}

\paragraph{Error-Driven Augmentation.}
A critical challenge in repository-level coding is robustness to distracting context. Standard SFT trains on the ``happy path'' of perfect localisation. However, in real-world inference, retrieval is imperfect; without training to handle noise, LLMs are prone to \textit{over-editing}: mistakenly modifying irrelevant files or unchanged regions because they are provided in context~\cite{zeng2025hyperedit}.
To mitigate this, we employ an intermediate model (Qwen-2.5-Coder-32B-Instruct) to generate realistic noise. 
For Fine-grained Navigation, we combine ground-truth files ($F_{gt}$) with erroneous hard negatives ($F_{neg}$) predicted by the intermediate model to define the training mapping:
$ \textit{Issue} + (F_{gt} \cup F_{neg}) \rightarrow \textit{Relevant Context}.$
Here, the model learns to extract \textit{Relevant Context} from $F_{gt}$ while returning ``No changes needed'' for $F_{neg}$.

Similarly, we harden the \textbf{Patch Generation (Step 3)} against noisy context. We utilise the intermediate model to retrieve distractor code regions ($C_{noise}$) that are semantically similar to the bug location but require no editing. By combining these with the correct content ($C_{relevant}$) to form the input \textit{Localised Context}, the training objective is defined as:
$ \textit{Issue} + (C_{relevant} \cup C_{noise}) \rightarrow \textit{Search/Replace}. $
This data-centric approach ensures the model discriminates based on the specific issue intent rather than merely operating on the assumption of perfect retrieval.

\begin{table}[t]
\caption{Statistics of the Agentless-Aligned SFT dataset. \textit{Origin} refers to positive samples derived from source benchmarks (SWE-bench-Live, SWE-rebench, SWE-Gym). \textit{Enhanced} refers to hard negative samples generated via error-driven augmentation.}
\label{tab:sft_stats}
\centering

\resizebox{\linewidth}{!}{
\begin{tabular}{lrrrr}
\toprule
\textbf{Data Source} & \textbf{Step 1} & \textbf{Step 2} & \textbf{Step 3} & \textbf{Total} \\
\midrule
Origin & 18,891 & 17,763 & 16,564 & 53,218 \\
Error-Augmented & -- & 12,989 & 8,875 & 21,864 \\
\midrule
\textbf{Total SFT Data} & \textbf{18,891} & \textbf{30,752} & \textbf{25,439} & \textbf{75,082} \\
\bottomrule
\end{tabular}
}

\end{table}

Table~\ref{tab:sft_stats} summarises the final composition of the SFT dataset. The \textit{Origin} set represents the clean trajectories derived from the source benchmarks, while the \textit{Error-Augmented} set comprises the negative samples generated via our error-driven augmentation pipeline to improve robustness against noise.

\section{Experiments}
\label{sec:experiments}

\begin{table*}[t]
\caption{Main results on SWE-bench Lite and Verified. The table is organised by benchmark split. All reported metrics are percentages. \textbf{File Acc.} denotes file-level localisation recall. \textbf{Line Acc.} denotes fine-grained navigation accuracy. \textbf{Valid} indicates the percentage of generated patches that are syntactically valid and can be applied successfully. \textbf{Pass@1} reports the percentage of issues resolved.}
\label{tab:main_results}
\centering
\resizebox{0.88 \linewidth}{!}{
\begin{tabular}{llccccc}
\toprule
\textbf{Base Model} & \textbf{Mid-Train Setting} & \textbf{SFT} & \textbf{Valid Patch} & \textbf{File Acc.} & \textbf{Line Acc.} & \textbf{Pass@1} \\
\midrule
\rowcolor[RGB]{234, 234, 234} \multicolumn{7}{c}{\textit{SWE-Bench Lite (300 instances)}} \\
\midrule
Qwen-Coder-32B-Instruct & None & \xmark & 77.0 & 74.7 & 38.3 & 10.7 \\
Qwen-Coder-32B-Base & None & \cmark & 84.0 & 78.3 & 46.7 & 11.3 \\
Qwen-Coder-32B-Base & StarCoder2-Style Dataset (All, 17.4B) & \cmark & 89.7 & 84.3 & 47.0 & 15.7 \\
\midrule
\multirow{2}{*}{Qwen-Coder-32B-Base} & \textbf{\model-train (Ours, Python, 3.8B)} & \cmark & 95.7 & 86.3 & 54.0 & 22.3 \\
& \textbf{\model-train (Ours, All, 17.7B)} & \cmark & \textbf{96.3} & \textbf{87.3} & \textbf{55.7} & \textbf{24.3} \\
\midrule
\rowcolor[RGB]{234, 234, 234} \multicolumn{7}{c}{\textit{SWE-Bench Verified (500 instances)}} \\
\midrule
Qwen-Coder-32B-Instruct & None & \xmark & 77.6 & 70.6 & 42.3 & 18.3 \\
Qwen-Coder-32B-Base & None & \cmark & 81.8 & 74.3 & 46.6 & 17.6 \\
Qwen-Coder-32B-Base & StarCoder2-Style Data (All, 17.4B) & \cmark & 82.4 & 77.7 & 48.4 & 20.4 \\
\midrule
\multirow{2}{*}{Qwen-Coder-32B-Base} & \textbf{\model-train (Ours, Python, 3.8B)} & \cmark & 94.4 & 78.5 & 51.6 & 27.8 \\
& \textbf{\model-train (Ours, All, 17.7B)} & \cmark & \textbf{95.2} & \textbf{80.7} & \textbf{52.2} & \textbf{30.6} \\
\bottomrule
\end{tabular}
}
\end{table*}

\subsection{Experiment Setup}

\paragraph{Training Configurations.}
We initialise our mid-training from Qwen2.5-Coder-32B-Base~\cite{hui2024qwen25codertechnicalreport} and conduct all experiments on a cluster of 32 NVIDIA H200 GPUs with a context window of 32,768 tokens.
For ablation analysis, we define a ``Python Only'' setting trained exclusively on the Python subset of \model-train, contrasting it with the full multi-language corpus.
In terms of computational cost, this ``Python Only'' mid-training requires approximately 60 wall-clock hours, significantly less than the 259 hours for the full ``All Languages'' setting, while the final stepwise SFT stage completes in 38 hours.
Comprehensive hyperparameter settings are provided in Appendix~\ref{app:training_details}.

\paragraph{Benchmarks and Metrics.}
We evaluate \model on SWE-bench Lite (300 instances) and Verified (500 instances) \citep{jimenez2024swebench}. 
We report four key metrics: (1) \textbf{Pass@1}, the primary metric for issue resolution; (2) \textbf{Valid Patch Rate}, measuring the percentage of generated patches that are applied successfully; and (3) intermediate retrieval metrics including \textbf{File localisation Accuracy} and \textbf{Line Accuracy}, which precisely quantify the model's ability to locate correct files and edit spans, respectively.

\paragraph{Inference Scaffold.} Crucially, we adopt a \textbf{Simplified Agentless} scaffolding \citep{xia2024agentless} (detailed in Section~\ref{subsec:sft}) rather than a complex Agent-based framework.
We choose this deterministic protocol for two reasons:
(1) \textbf{Lightweight Evaluation:} The Agentless workflow encapsulates the standard problem-solving stages (localisation,  patch generation) found in most agentic frameworks but executes them in a linear, efficient manner. This avoids the heavy computational overhead of iterative execution loops, enabling rapid and scalable benchmarking. 
(2) \textbf{Isolation of Gains:} This streamlined workflow allows us to \textbf{more clearly and reliably} isolate and measure the intrinsic editing capabilities acquired from our data pipeline, disentangling our contribution from the variance introduced by complex planning loops or prompt engineering strategies.

\paragraph{Internal Baselines: Data Strategy Ablation.}
We compare \model against three controlled settings based on Qwen2.5-Coder-32B.
First, we use its official Instruct model as a zero-shot baseline to represent generalist capabilities.
Second, we evaluate Base + SFT (without mid-training) to establish a lower bound for instruction tuning.
Third, and most critically, we implement a StarCoder2-style Baseline \citep{lozhkov2024starcoder2stackv2}. This baseline represents the prevailing standard for training on GitHub data (Appendix~\ref{app:starcoder2_baseline}) but differs from \model in three fundamental aspects:
(1) \textbf{Format:} It utilises the noisy \textit{Unified Diff} format rather than our verifiable Search/Replace blocks;
(2) \textbf{Filtering:} It retains non-code artefacts (e.g., JSON, YAML config files), whereas we enforce strict Core Language filtering; and
(3) \textbf{Issue-Augmented Context:} Unlike standard practice which processes PRs and Issues in isolation, \model integrates linked Issue descriptions into the training sequence. This forces the model to learn the alignment between natural language intent and code implementation.

\paragraph{External Baselines: Open-Source SOTA.}
We further benchmark \model against representative high-performing open systems to contextualise our efficiency. We include SWE-Gym \citep{pan2025swegym}, which shares our model size (32B) but employs the complex OpenHands agentic framework with iterative planning. Additionally, we compare against substantially larger models, specifically Lingma-SWE \citep{ma2024lingmaswegptopendevelopmentprocesscentric} and SWE-Fixer \citep{xie2025swefixer}. Both utilise 72B-parameter base models.

\begin{table}[t]
\caption{Comparison with open-source methods (pass@1); results are copied from the original paper. All metrics are percentages.}
\label{tab:sota_comparison}
\centering
\resizebox{\linewidth}{!}{
\begin{small}
\begin{tabular}{llccc}
\toprule
\textbf{Method} & \textbf{Framework} & \textbf{Params} & \textbf{Lite} & \textbf{Verified} \\
\midrule
SWE-Gym  & OpenHands & 32B & 15.3 & 20.6 \\
Lingma-SWE  & SWESynInfer & \textbf{72B} & 22.0 & 30.2 \\
SWE-Fixer& SWE-Fixer & \textbf{72B} & 22.0 & 30.2 \\
\rowcolor[RGB]{234, 234, 234} \textbf{\model} & Agentless & 32B & \textbf{24.3} & \textbf{30.6} \\
\bottomrule
\end{tabular}
\end{small}
}
    \vskip -0.1in
\end{table}

\subsection{Main Results}
\label{subsec:main_results}

Table~\ref{tab:main_results} presents the evaluation on SWE-bench Lite and Verified. We analyse the results across three key dimensions: the effectiveness of mid-training, the impact of \model pipeline, and the benefits of multi-language training.

\paragraph{Effectiveness of Mid-Training.}

The results highlight the benefit of incorporating a dedicated repository-level mid-training stage.
The Base + SFT model, despite being fine-tuned on the stepwise dataset, achieves only 11.3\% on Lite and 17.6\% on Verified. In contrast, introducing any form of repository-level mid-training, even the noisy StarCoder2-style baseline, yields immediate gains. This step boosts Lite performance to 15.7\% (+4.4\%) and Verified to 20.4\% (+2.8\%), confirming that pre-encoding repository structures and editing patterns into the model weights is a prerequisite for effective downstream performance.

\paragraph{Superiority of \model Data Construction.}

Comparing StarCoder2-style (17.4B) with \model-train (17.7B) shows a clear advantage, with our method reaching 24.3\% on Lite and 30.6\% on Verified.
We hypothesize these gains stem from our strict data structuring.
First, the verified \textbf{Search/Replace} format correlates with improved \textit{Valid Patch} rates (89.7\% $\to$ 96.3\%). Unlike line-number-based Diffs, S/R requires explicit context matching, which likely grounds the model's edits and reduces format application errors.
Second, the rise in \textit{Line Localisation} (47.0\% $\to$ 55.7\%) suggests that training on unique search blocks encourages the model to generate more precise, unambiguous navigation cues compared to raw noisy diffs.

\paragraph{Benefits of Multi-Language Training.}
We find that multi-language training yields additional performance gains.
Although the Python Only model (3.8B tokens) performs impressively, surpassing the 17.4B token StarCoder baseline with 22.3\% on Lite, scaling to All Languages (17.7B tokens) yields the best overall results (24.3\% on Lite and 30.6\% on Verified). This suggests that exposure to diverse syntactical structures, such as those from Java, C++, and Go, enhances the abstract reasoning capabilities of the model in issue solving/software engineering.

\paragraph{Comparison with Recent Open-Source methods.}
Table~\ref{tab:sota_comparison} benchmarks \model against representative open-source methods. Using the same 32B base, \model significantly outperforms SWE-Gym (30.6\% vs 20.6\% on Verified) which relies on complex agent scaffolding. Remarkably, despite having half the parameters, our model surpasses the 72B baselines (Lingma-SWE and SWE-Fixer) on both Lite and Verified. This confirms that rigorous mid-training bridges the scaling gap, enabling SOTA performance under a lightweight workflow without expensive iterative loops.

\subsection{Ablation Studies}
\label{subsec:ablation}

\begin{table}[t]
\caption{Effect of data source and edit format on SWE-Bench performance. We bold our default setting for comparison.}
\label{tab:data_source_edit_format}
\centering
\resizebox{\linewidth}{!}{
\begin{tabular}{lllc c}
\toprule
\textbf{Data source} & \textbf{Edit Format} & \textbf{Description} & \textbf{Lite} & \textbf{Verified} \\
\midrule
StarCoder-style & Diff & PR Desc Only & 15.7 & 20.4 \\
\midrule
\multirow{3}{*}{\shortstack{\model-train\\(Python)}} & \textbf{Search/Replace} & \textbf{Linked Issue} & \textbf{22.3} & \textbf{27.8} \\
& Diff & Linked Issue & 19.1 & 24.4 \\
& Search/Replace & PR Desc Only & 20.4 & 25.7 \\
\bottomrule
\end{tabular}
}
\end{table}

\paragraph{Contributions of Linked Issue and Edit Format.}
To rigorously disentangle the individual contributions of our data construction pipeline, we conduct an ablation study on the Python subset of our mid-training data, as detailed in Table~\ref{tab:data_source_edit_format}, with two main findings.
\textbf{First, the edit format is critical.} Replacing our verified Search/Replace blocks with standard Unified Diffs results in a performance drop (e.g., from 27.8\% down to 24.4\% on Verified). This confirms that the deterministic, context-rich nature of Search/Replace blocks provides a far more robust training signal than brittle diff lines.
\textbf{Second, augmenting context provides a realistic problem definition.} 
Relying solely on raw PR descriptions degrades performance to 25.7\% on Verified. By incorporating linked issue descriptions, we provide the model with the original problem definition rather than just the solution summary, yielding a clear gain.
Notably, the combination of both strategies achieves the best performance, significantly outperforming the StarCoder-style baseline which lacks both rigorous formatting and intent augmentation.

\begin{table}[h]
\caption{Ablation study of SFT data strategies on SWE-Bench Lite and Verified.}
\label{tab:ablation_sft_full}
\centering
\resizebox{\linewidth}{!}{
\begin{tabular}{l l ccc ccc}
\toprule
\multirow{2}{*}{\textbf{Lang}} & \multirow{2}{*}{\makecell[l]{\textbf{SFT}\\\textbf{Strategy}}} &
\multicolumn{3}{c}{\textbf{Lite}} & \multicolumn{3}{c}{\textbf{Verified}} \\
\cmidrule(lr){3-5}\cmidrule(lr){6-8}
& & \textbf{File} & \textbf{Line} & \textbf{Pass@1} & \textbf{File} & \textbf{Line} & \textbf{Pass@1} \\
\midrule
\multirow{2}{*}{Python} & Standard & 86.7 & 51.3 & 18.7 & 78.3 & 49.4 & 24.3 \\
& \textbf{Error Aug.} & \textbf{86.3} & \textbf{54.0} & \textbf{22.3} & \textbf{78.5} & \textbf{51.6} & \textbf{27.8} \\
\midrule
\multirow{2}{*}{All} & Standard & 87.0 & 53.0 & 21.8 & 80.3 & 50.0 & 27.4 \\
& \textbf{Error Aug.} & \textbf{87.3} & \textbf{55.7} & \textbf{24.3} & \textbf{80.7} & \textbf{52.2} & \textbf{30.6} \\
\bottomrule
\end{tabular}
}
\end{table}

\textbf{Impact of Error-Driven Augmentation.}
To validate the effectiveness of our augmentation strategy, we compare the performance of models fine-tuned on the standard SFT dataset versus the version augmented with hard negatives and distractor regions. As shown in Table~\ref{tab:ablation_sft_full}, this strategy yields consistent gains across all settings. For our best-performing ``All Languages'' model, the augmentation boosts the Pass@1 rate from 21.8\% to \textbf{24.3\%} on SWE-bench Lite and from 27.4\% to \textbf{30.6\%} on SWE-bench Verified. Crucially, we observe simultaneous improvements in Line accuracy, which confirms that explicitly training the model to discriminate against distracting context and reject irrelevant files significantly enhances its robustness and precision in real-world repository navigation.

\begin{figure}[t]
\centering
\includegraphics[width=0.95\columnwidth]{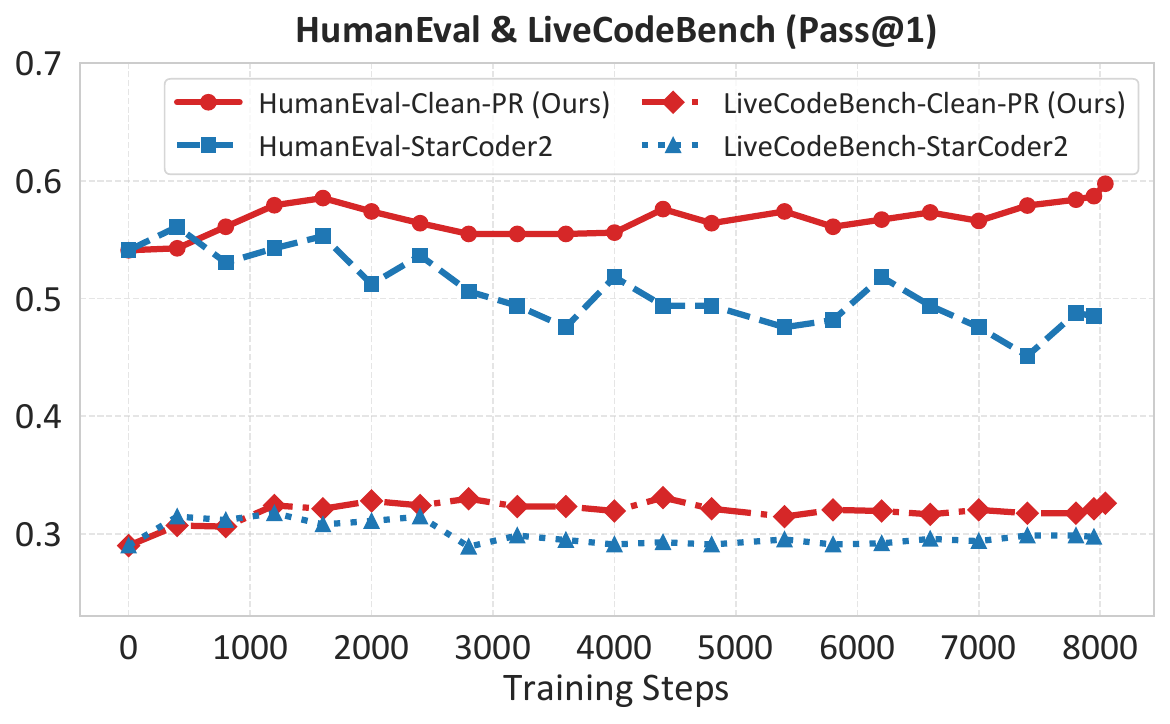}
\caption{Generalisation dynamics during mid-training.}
\label{fig:generalization_curves}
\vskip -0.1in
\end{figure}

\paragraph{Generalisation Capability and Catastrophic Forgetting.}
A critical challenge in repository-specific adaptation is avoiding the loss of general programming capabilities (``catastrophic forgetting'')~\cite{van_de_Ven_2025}. We visualise the training dynamics in Figure~\ref{fig:generalization_curves}.
The StarCoder2-style baseline, trained on standard diffs, exhibits a clear degradation trend: HumanEval performance drops from 54.1\% to 47.6\% (-6.5\%) as training progresses. This suggests that raw diffs may hinder the model's core reasoning due to fragile line numbers and unverified context.
In stark contrast, \model demonstrates robust positive transfer. By learning from \textbf{verified Search/Replace} blocks, the model not only preserves its pre-trained capabilities but actively sharpens them, reaching 59.8\% on HumanEval (+5.7\%) and boosting LiveCodeBench from 29.0\% to 32.6\%. 
This suggests that the precise context matching required by our objective transfers effectively to general code generation, proving that repository-level adaptation need not come at the cost of fundamental coding skills.

\begin{table*}[t]
\caption{Scale generalisation on Qwen2.5-Coder-7B. The table is organised by benchmark split. All reported metrics are percentages. \textbf{File Acc.} denotes file-level localisation recall. \textbf{Line Acc.} denotes fine-grained navigation accuracy. \textbf{Valid} indicates the percentage of generated patches that are syntactically valid and can be applied successfully. \textbf{Pass@1} reports the percentage of issues resolved.}
\label{tab:qwen7b_results}
\centering
\resizebox{0.88\linewidth}{!}{
\begin{tabular}{llccccc}
\toprule
\textbf{Base Model} & \textbf{Mid-Train} & \textbf{SFT} & \textbf{Valid Patch} & \textbf{File Acc.} & \textbf{Line Acc.} & \textbf{Pass@1} \\
\midrule
\rowcolor[RGB]{234, 234, 234} \multicolumn{7}{c}{\textit{SWE-Bench Lite (300 instances)}} \\
\midrule
SWE-Gym (Qwen2.5-Coder-7B-Instruct) & None & \cmark & -- & -- & -- & 10.0 \\
Lingma-SWE-GPT-7B & External & -- & -- & -- & -- & 12.0 \\
Qwen-Coder-7B-Instruct & None & \xmark & 6.3 & 62.7 & 24.4 & 1.3 \\
Qwen-Coder-7B-Base & None & \cmark & 81.2 & 66.3 & 32.0 & 10.3 \\
Qwen-Coder-7B-Base & \textbf{\model} & \cmark & \textbf{90.7} & \textbf{82.0} & \textbf{42.7} & \textbf{14.5} \\
\midrule
\rowcolor[RGB]{234, 234, 234} \multicolumn{7}{c}{\textit{SWE-Bench Verified (500 instances)}} \\
\midrule
SWE-Gym (Qwen2.5-Coder-7B-Instruct) & None & \cmark & -- & -- & -- & 10.6 \\
Lingma-SWE-GPT-7B & External & -- & -- & -- & -- & 18.2 \\
Qwen-Coder-7B-Instruct & None & \xmark & 5.7 & 57.6 & 22.6 & 2.4 \\
Qwen-Coder-7B-Base & None & \cmark & 80.1 & 61.3 & 29.5 & 14.2 \\
Qwen-Coder-7B-Base & \textbf{\model} & \cmark & \textbf{90.8} & \textbf{77.0} & \textbf{43.2} & \textbf{20.4} \\
\bottomrule
\end{tabular}
}
\end{table*}

\begin{figure}[t]
\centering
\includegraphics[width=1 \columnwidth]{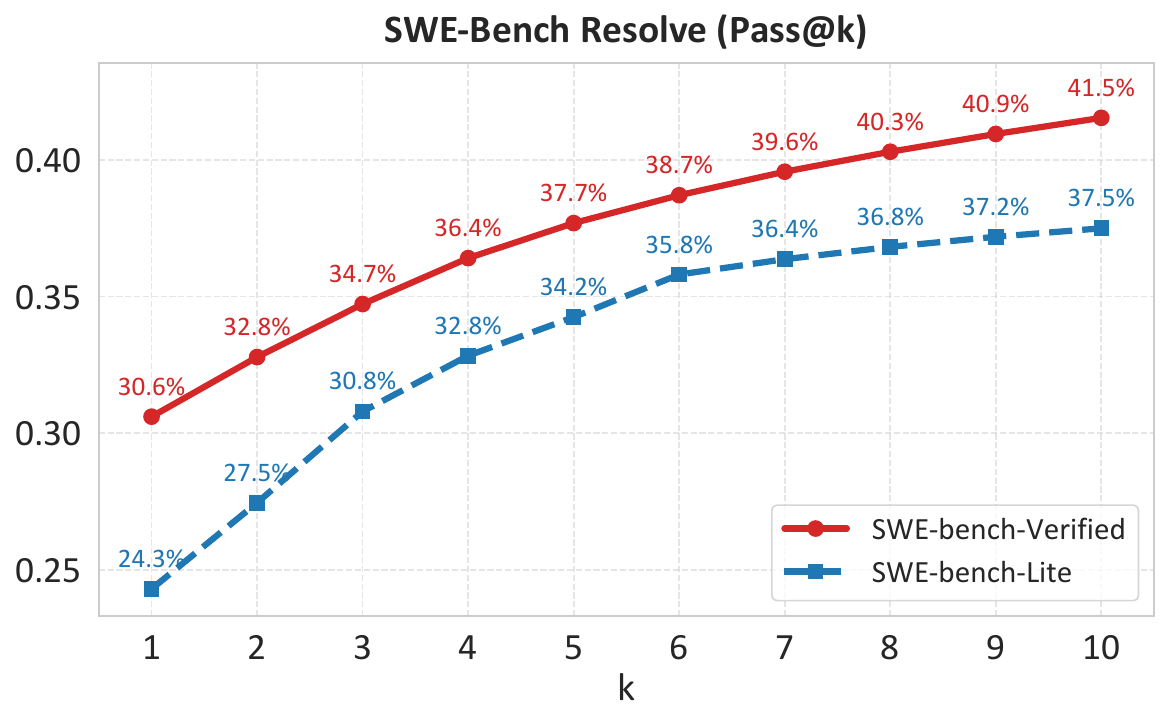}
\caption{Pass@k performance on SWE-bench Lite and Verified. We report the resolution rates of our model (\model, mid-trained on All Languages) as the number of samples $k$ scales.}
\label{fig:pass_k_curve}
\vskip -0.1in
\end{figure}

\paragraph{Scaling Inference with Best-of-N.}
\label{sec:pass@k_part}
We explore the upper bound of our model's capability by evaluating Pass@k performance, where the model generates $k$  candidate patches for each issue. As illustrated in Figure~\ref{fig:pass_k_curve}, \model benefits substantially from increased sampling.
On SWE-bench Verified, the performance improves monotonically from 30.6\% at $k=1$ to 41.5\% at $k=10$. Similarly, on SWE-bench Lite, the resolution rate rises from 24.3\% to 37.5\%.
This gap between Pass@1 and Pass@10 suggests that while the model has the intrinsic reasoning capability to solve a large portion of issues, the standard likelihood-based ranking is not always perfectly aligned with functional correctness. These results indicate that integrating a lightweight re-ranking mechanism or a verifier could further unlock the model's potential without requiring expensive agentic training.

\subsection{Additional Generalisation Studies}
\label{subsec:additional_generalization}


\paragraph{Scale Generalisation.}
We further investigate whether the benefits of \model generalise across model scales beyond the default 32B setting.
Specifically, we apply the same mid-training and SFT pipeline to Qwen2.5-Coder-7B and compare against both internal and external 7B baselines (SWE-Gym fine-tuned on Qwen2.5-Coder-7B-Instruct, and Lingma-SWE-GPT-7B).
Table~\ref{tab:qwen7b_results} shows that the same recipe consistently transfers to Qwen2.5-Coder-7B, improving Pass@1 from 10.3\% to 14.5\% on Lite and from 14.2\% to 20.4\% on Verified over the SFT-only base.
Notably, the localisation gains are larger than for the 32B model, suggesting that high-quality PR supervision is especially useful when model capacity is limited.


\paragraph{Multilingual Evaluation.}
The main SWE-bench splits are Python-only, whereas \model-train spans 12 languages. Table~\ref{tab:multiswe_results} therefore evaluates multilingual issue-solving on Multi-SWE-bench Flash~\citep{zan2025multiswebench}, which contains 300 instances across C, C++, Go, Java, JavaScript, Rust, and TypeScript. For multilingual SFT, we use Multi-SWE-bench-RL, yielding 9,331 samples from 69 repositories with no overlap with SWE-bench Lite/Verified or Multi-SWE-bench. \model reaches 12.3\% Pass@1, outperforming both the instruct baseline and the StarCoder2-style baseline, with consistent gains in valid-patch and localisation.

\begin{table}[t]
\caption{Multilingual evaluation on Multi-SWE-bench Flash. All metrics are percentages.}
\label{tab:multiswe_results}
\centering
\resizebox{\linewidth}{!}{
\begin{tabular}{llccccc}
\toprule
\textbf{Base Model} & \textbf{Mid-Train} & \textbf{SFT} & \textbf{Valid} & \textbf{File} & \textbf{Line} & \textbf{Pass@1} \\
\midrule
Qwen-Coder-32B-Instruct & None & \xmark & 71.7 & 40.0 & 15.2 & 6.7 \\
Qwen-Coder-32B-Base & None & \cmark & 73.0 & 42.3 & 17.7 & 7.0 \\
Qwen-Coder-32B-Base & StarCoder2 & \cmark & 76.3 & 46.0 & 20.3 & 8.7 \\
Qwen-Coder-32B-Base & \textbf{\model} & \cmark & \textbf{81.7} & \textbf{51.3} & \textbf{24.0} & \textbf{12.3} \\
\bottomrule
\end{tabular}
}
\end{table}

\paragraph{Agentic Evaluation.}
Finally, we test whether the internalised editing capability transfers back into a multi-turn agent. Following SWE-Gym, we train an agent variant with 491 successful OpenHands trajectories~\citep{wang2024openhands} and evaluate on OpenHands v0.28 using CodeActAgent with a 100-turn limit. As shown in Table~\ref{tab:openhands_results}, \model improves over SFT-only by 4.6 points on Lite and 5.2 points on Verified, while also reducing empty-patch rates. Since OpenHands uses a different action space and failure modes from our Simplified Agentless scaffold, this suggests that \model provides a stronger editing prior for agents rather than merely matching a fixed inference template.

\begin{table}[t]
\caption{Agentic evaluation with OpenHands CodeActAgent.}
\label{tab:openhands_results}
\centering
\resizebox{\linewidth}{!}{
\begin{tabular}{lcccc}
\toprule
\textbf{Model} & \textbf{Empty} $\downarrow$ & \textbf{Loop} $\downarrow$ & \textbf{Turns} & \textbf{Resolve} $\uparrow$ \\
\midrule
\rowcolor[RGB]{234, 234, 234} \multicolumn{5}{c}{\textit{SWE-Bench Lite (300 instances)}} \\
\midrule
SWE-Gym & -- & -- & -- & 15.3 \\
Qwen-Coder-32B-Instruct & 42.3 & 42.8 & 29.3 & 2.8 \\
Qwen-Coder-32B-Base+SFT & 21.0 & 28.7 & 37.8 & 16.1 \\
Qwen-Coder-32B-Base+\textbf{\model}+SFT & \textbf{16.3} & \textbf{26.7} & 42.5 & \textbf{20.7} \\
\midrule
\rowcolor[RGB]{234, 234, 234} \multicolumn{5}{c}{\textit{SWE-Bench Verified (500 instances)}} \\
\midrule
SWE-Gym & -- & -- & -- & 20.6 \\
Qwen-Coder-32B-Instruct & 28.4 & 37.8 & 27.4 & 6.3 \\
Qwen-Coder-32B-Base+SFT & 16.6 & \textbf{23.7} & 34.1 & 19.5 \\
Qwen-Coder-32B-Base+\textbf{\model}+SFT & \textbf{14.3} & 24.7 & 38.4 & \textbf{24.7} \\
\bottomrule
\end{tabular}
}
\vskip -0.08in
\end{table}

\section{Related Work}

\paragraph{Inference Paradigms and System Complexity.}
The pursuit of automated repository-level engineering has spurred a diverse ecosystem of inference frameworks. 
Early dominant approaches relied on \textbf{Agentic frameworks}, where models function as autonomous agents interacting with an environment via tools (e.g., shell, file editors). Systems like SWE-agent \citep{yang2024sweagent}, OpenHands \citep{wang2024openhands}, and AutoCodeRover \citep{zhang2024autocoderover} employ iterative reasoning loops to navigate codebases, though they often suffer from error propagation in long trajectories. 
In response, \textbf{Agentless paradigms} \citep{xia2024agentless} emerged as a streamlined alternative, decomposing the problem into static retrieval, precise localisation, recently enhanced by code-structure signals like call graphs \citep{jiang2025cosil}, and constrained patch synthesis. 
However, the recent trend has shifted towards overcoming model limitations through increasing system complexity and \textbf{Test-Time Scaling}. 
This includes the development of reinforcement learning environments \citep{pan2025swegym,jain2025r2egym} for policy optimisation, contamination-aware evaluation protocols \citep{badertdinov2025swerbench}, and compute-intensive search strategies that sample and rerank candidate trajectories \citep{antoniades2024swesearch}. 
Recent systems such as SWE-Swiss-32B~\citep{sweswiss2025} further combine multi-task SFT, reinforcement learning, and self-consistency decoding, which are complementary to our focus on data-centric mid-training.
Furthermore, benchmarks have evolved to challenge these systems with dynamic issue streams \citep{zhang2025swebenchlive}, multilingual repositories \citep{zan2025multiswebench,rashid2025swepolybench}, and long-context understanding \citep{rando2025longcodebench}. 
While these engineering advancements drive higher scores, they often obscure the intrinsic capability of the underlying model. Our work focuses on internalising these repository-editing skills directly into model weights, reducing the dependency on heavy inference scaffolding.

\paragraph{Evolution of Code Training: From Files to PRs.}
The efficacy of code models is fundamentally constrained by the granularity of their training data, which has evolved through three distinct levels.
\textbf{1) File-Level:} Foundational models like StarCoder \citep{lozhkov2024starcoder2stackv2} and Qwen-Coder\cite{hui2024qwen25codertechnicalreport} are pre-trained on massive file collections such as The Stack \citep{kocetkov2022stack}. While this provides vast syntactic knowledge, it treats code as static snapshots, lacking the temporal context of software evolution. \textbf{2) Commit-Level:} To capture editing dynamics, recent work leverages version-control diffs and commit metadata, ranging from instruction tuning on commits (e.g., CommitPackFT \citep{muennighoff2024octopack}, Commitbench~\cite{schall2024commitbenchbenchmarkcommitmessage}) to commit- and edit-centric pretraining objectives (e.g., CoditT5 \citep{zhang2022coditt5}, CommitBART \citep{liu2022commitbart}, Coeditor \citep{wei2024coeditor}). 
However, commits/diffs still provide weak or fragmented intent signals \citep{tian2022goodcommitmessage}, and rarely capture the full multi-file context and discussion that drive real engineering work.
\textbf{3) PR-Level:} Pull Requests represent the ideal training signal, offering a comprehensive view that couples high-level human intent with extensive, multi-file code modifications \citep{10.1145/2568225.2568260,tsay2014influence}. 
Despite their potential, leveraging PRs is notoriously difficult due to the ``noise-validity gap" in mining GitHub at scale, further exacerbated by PR-specific artefacts such as bot-generated activity \citep{golzadeh2020botornot,wessel2020inconvenient}, and the prevalence of unmerged or abandoned contributions that lack verifiable quality assurance.
Consequently, prior PR-centric works have been limited to auxiliary tasks like code review \citep{li2022automatingcodereviewactivities} or synthetic data generation \citep{wei2023magicoder,yang2025swesmith}, rather than direct training for editing. 
We bridge this gap by proposing \textbf{\model}, a scalable pipeline that rigorously filters and verifies PRs to construct a massive, deterministic corpus, enabling models to learn repository-level editing at scale.

\section{Conclusion}
\label{sec:conclusion}

In this work, we addressed the scarcity of high-quality supervision for repository-level engineering by introducing \textbf{\model}. We transformed noisy GitHub pull requests into a rigorous corpus of \textbf{2 million} verifiable \textbf{Search/Replace} instances (17.7B tokens). To harness this data, we proposed an \textbf{Agentless-aligned stepwise SFT} strategy augmented with error-driven negative sampling. Our experiments demonstrate that this data-centric approach enables a 32B model to achieve highly competitive performance without heavy agent scaffolding, while additional 7B, multilingual, and OpenHands evaluations show that the learned repository-editing capability transfers across scale, language, and inference paradigms.



\section*{Impact Statement}

This work introduces \model\ to advance automated repository-level software engineering. While our approach holds significant potential to enhance developer productivity and lower the barrier for open-source maintenance, it also introduces challenges regarding the reliability and security of generated code. As models gain the capability to modify complex codebases, there are risks of introducing subtle vulnerabilities or being misused for malicious automation. Furthermore, the use of public repositories necessitates ongoing attention to licensing and attribution. We encourage the community to prioritise the development of robust verification tools and safety guardrails to ensure these agents serve as reliable, human-aligned assistants.

\section*{Acknowledgments}
This work was supported in part by the UK Engineering and Physical Sciences Research Council (EPSRC) through a Turing AI Fellowship (grant no. EP/V020579/1, EP/V020579/2) and a New Horizons grant (grant no. EP/X019063/1), and KCL’s Impact Acceleration Account (grant no. EP/X525571/1).  The work of Xiangxiang Dai was supported by the National Natural Science Foundation of China (625B2163).


\bibliography{example_paper}
\bibliographystyle{icml2026}

\newpage
\appendix
\clearpage

\onecolumn
\newpage
\addtocontents{toc}{\protect\setcounter{tocdepth}{3}}
\renewcommand{\contentsname}{Appendix}

{
\hypersetup{linkcolor=black}
\tableofcontents %

}
\newpage
\appendix

\twocolumn
\newpage

\begin{table*}[t]
    \centering
    \caption{\textbf{Visualisation of Raw Data Entities.} Note that Issues and PRs are distinct objects. The Issue contains the natural language description of the bug, while the PR contains the metadata, code context, and the diff. We link them during the data construction phase.}
    \label{tab:raw_data_instances}
    
    \begin{tcolorbox}[
        width=\textwidth, 
        colback=orange!5!white, 
        colframe=orange!60!black, 
        title={\textbf{\textsc{Data Instance 1}: Raw Issue \#4288 (User Intent)}},
        fonttitle=\bfseries,
        sharp corners=south
    ]
        \textbf{Repo:} \texttt{example/repository} \hfill \textbf{Issue ID:} \texttt{4288} \hfill \textbf{Author:} \texttt{@user\_A}\\
        \textbf{Created At:} 2024-03-14 10:00:00 \hfill \textbf{Status:} \texttt{Closed}
        
        \rule{\linewidth}{0.4pt}
        
        \textbf{Title:} \texttt{IndexError when batch size is 1 in attention mask}\\   
        \textbf{Description Body:} \\
        When passing a single sequence to the model (batch size = 1), the attention mask expansion fails.
        \begin{verbatim}
Traceback (most recent call last):
  File "src/models/attention.py", line 45, in expand_mask
    bsz, src_len = mask.size()
ValueError: not enough values to unpack (expected 2, got 1)        \end{verbatim}
        It seems we strictly expect a batch dimension $>1$. This works fine when $bsz>1$ but crashes on single inference.
    \end{tcolorbox}

    \vspace{0.2cm} 

    \begin{tcolorbox}[
        width=\textwidth, 
        colback=blue!3!white, 
        colframe=blue!50!black, 
        title={\textbf{\textsc{Data Instance 2}: Raw PR \#4290 (Implementation)}},
        fonttitle=\bfseries,
        sharp corners=north
    ]
        \textbf{Repo:} \texttt{example/repository} \hfill \textbf{PR ID:} \texttt{4290} \hfill \textbf{Author:} \texttt{@dev\_B}\\
        \textbf{Status:} \texttt{Merged} \hfill \textbf{Base Commit:} \texttt{7b3f1a2} \hfill \textbf{Head Commit:} \texttt{9c4e2d1}
        
        \rule{\linewidth}{0.4pt}
        \textbf{PR Title:} \texttt{Fix crash on empty input list} \\
        \textbf{PR Description:} \texttt{Fixed the unpacking error reported in \#4288. Added specific shape check.}
        
        \vspace{0.2cm}
        \textbf{File Context (Base Code before edit):} \hfill \textit{*Path: src/models/attention.py}
        \begin{tcolorbox}[colback=gray!10!white, colframe=gray!50!white, boxsep=0pt, left=2pt, right=2pt, top=2pt, bottom=2pt]
\begin{verbatim}
38  def expand_mask(mask, dtype, tgt_len):
39      """
40      Expands attention_mask to [bsz, 1, tgt_len, src_len].
41      """
42      # CAUTION: This line causes overflow if not careful
43      bsz, src_len = mask.size() 
44      tgt_len = tgt_len if tgt_len is not None else src_len
45
46      expanded_mask=mask[:,None,None,:].expand(bsz,1,tgt_len,src_len)
47      return expanded_mask.to(dtype)
\end{verbatim}
        \end{tcolorbox}

        \textbf{Diff Hunk (The Change):}
        \begin{tcolorbox}[colback=green!5!white, colframe=green!40!black, boxsep=0pt, left=2pt, right=2pt, top=2pt, bottom=2pt]
\begin{verbatim}
diff --git a/src/models/attention.py b/src/models/attention.py
--- a/src/models/attention.py
+++ b/src/models/attention.py
@@ -43,3 +43,5 @@ def expand_mask(mask, dtype, tgt_len):
-     bsz, src_len = mask.size()
+     # Robust unpacking for single batch
+     shape = mask.shape
+     bsz, src_len = shape[0], shape[-1]
      tgt_len = tgt_len if tgt_len is not None else src_len
\end{verbatim}
        \end{tcolorbox}
        
        \textbf{Comments:} \\
        \texttt{[@maintainer]: Verified. The .shape access is safer here. Merging.}
    \end{tcolorbox}
\end{table*}

\clearpage

\clearpage
\begin{table*}[h]
    \centering
    \caption{\textbf{Why Convert? Diff vs. Search/Replace.} (Left) Raw Git Diffs use line numbers (e.g., \texttt{@@ -43,3}), which makes them fragile for training; if the code shifts by one line, the label becomes invalid. (Right)  Search/Replace format uses unique context strings to anchor the edit, ensuring robustness.}
    \label{tab:diff_vs_sr}
    \small
    \begin{tabular*}{\textwidth}{p{0.48\textwidth} p{0.48\textwidth}}
        \toprule
        \multicolumn{1}{c}{\textbf{Raw Format: Unified Diff}} & \multicolumn{1}{c}{\textbf{Training Format: Search/Replace}} \\
        \midrule
        \begin{tcolorbox}[colback=red!5!white, colframe=red!50!black, arc=0mm, left=1pt, right=1pt]
\begin{verbatim}
diff --git a/attention.py b/attention.py
--- a/attention.py
+++ b/attention.py
@@ -43,3 +43,5 @@  <-- FRAGILE
-     bsz, src_len = mask.size()
+     shape = mask.shape
+     bsz, src_len = shape[0], shape[-1]
     tgt_len = tgt_len...
\end{verbatim}
        \end{tcolorbox} 
        \textbf{Cons:} Relies on \texttt{line 43}. If upstream changes move this to \texttt{line 45}, the patch fails.
        & 
        \begin{tcolorbox}[colback=green!5!white, colframe=green!50!black, arc=0mm, left=1pt, right=1pt]
\begin{verbatim}
<<<<<<< SEARCH
     bsz, src_len = mask.size()
     tgt_len = tgt_len if tgt_len
=======
     shape = mask.shape
     bsz, src_len = shape[0], shape[-1]
     tgt_len = tgt_len if tgt_len
>>>>>>> REPLACE
\end{verbatim}
        \end{tcolorbox}
        \textbf{Pros:} Matches the text \texttt{bsz, src\_len...} anywhere in the file.
        \\
        \bottomrule
    \end{tabular*}
\end{table*}

\begin{table*}[h]
\caption{Language definitions used for data filtering. A PR is assigned to a language $L$ if it modifies at least one \textbf{Core} file of $L$, and contains only files from the \textbf{Allowed} list of $L$.}
\label{tab:language_rules}
\centering
\resizebox{1\linewidth}{!}{
\begin{tabular}{l p{0.15\linewidth} p{0.65\linewidth}}
\toprule
\textbf{Language} & \textbf{Core Extensions} & \textbf{Allowed Extensions (Context \& Config)} \\
\midrule
Python & .py & .py, .md, .rst, .txt, .yml, .yaml, .toml, .cfg, .ini, .json, .png, .jpg, .jpeg, .svg, .gif, .html, .sh, .bash \\
\midrule
Java & .java & .java, .xml, .properties, .gradle, .md, .txt, .json, .yml, .yaml, .png, .jpg, .jpeg, .svg, .gif, .html, .css, .js, .sh \\
\midrule
TypeScript & .ts, .tsx & .ts, .tsx, .js, .jsx, .json, .md, .txt, .yml, .yaml, .png, .jpg, .jpeg, .svg, .gif, .vue, .html, .css, .scss, .sass, .less, .sh, .graphql, .gql \\
\midrule
Go & .go & .go, .mod, .sum, .proto, .md, .txt, .yml, .yaml, .json, .png, .jpg, .jpeg, .svg, .gif, .html, .sh \\
\midrule
Kotlin & .kt, .kts & .kt, .kts, .java, .xml, .gradle, .properties, .md, .txt, .json, .yaml, .yml, .toml, .png, .jpg, .jpeg, .svg, .gif, .html, .sh \\
\midrule
JavaScript & .js, .jsx & .js, .jsx, .json, .md, .txt, .yml, .yaml, .vue, .png, .jpg, .jpeg, .svg, .gif, .html, .css, .scss, .sass, .less, .sh \\
\midrule
C++ & .cpp, .cc, .cxx, .c++, .hpp, .hh, .hxx & .cpp, .cc, .cxx, .c++, .hpp, .h, .hh, .hxx, .c, .cmake, .txt, .md, .json, .yml, .yaml, .mk, .png, .jpg, .jpeg, .svg, .gif, .html, .sh \\
\midrule
C & .c, .h & .c, .h, .cmake, .txt, .mk, .makefile, .md, .json, .yml, .yaml, .png, .jpg, .jpeg, .svg, .gif, .html, .sh \\
\midrule
Rust & .rs & .rs, .toml, .lock, .md, .txt, .png, .jpg, .jpeg, .svg, .gif, .html, .json, .sh \\
\midrule
Ruby & .rb & .rb, .erb, .rake, .gemspec, .yml, .yaml, .md, .txt, .png, .jpg, .jpeg, .svg, .gif, .html, .json, .sh \\
\midrule
PHP & .php & .php, .xml, .yml, .yaml, .ini, .md, .txt, .png, .jpg, .jpeg, .svg, .gif, .json, .html, .sh \\
\midrule
C\# & .cs & .cs, .csproj, .sln, .json, .xml, .config, .md, .txt, .png, .jpg, .jpeg, .svg, .gif, .html, .sh \\
\bottomrule
\end{tabular}
}
\end{table*}

\clearpage

\begin{figure*}[t]
\centering
\includegraphics[width=1.0\textwidth]{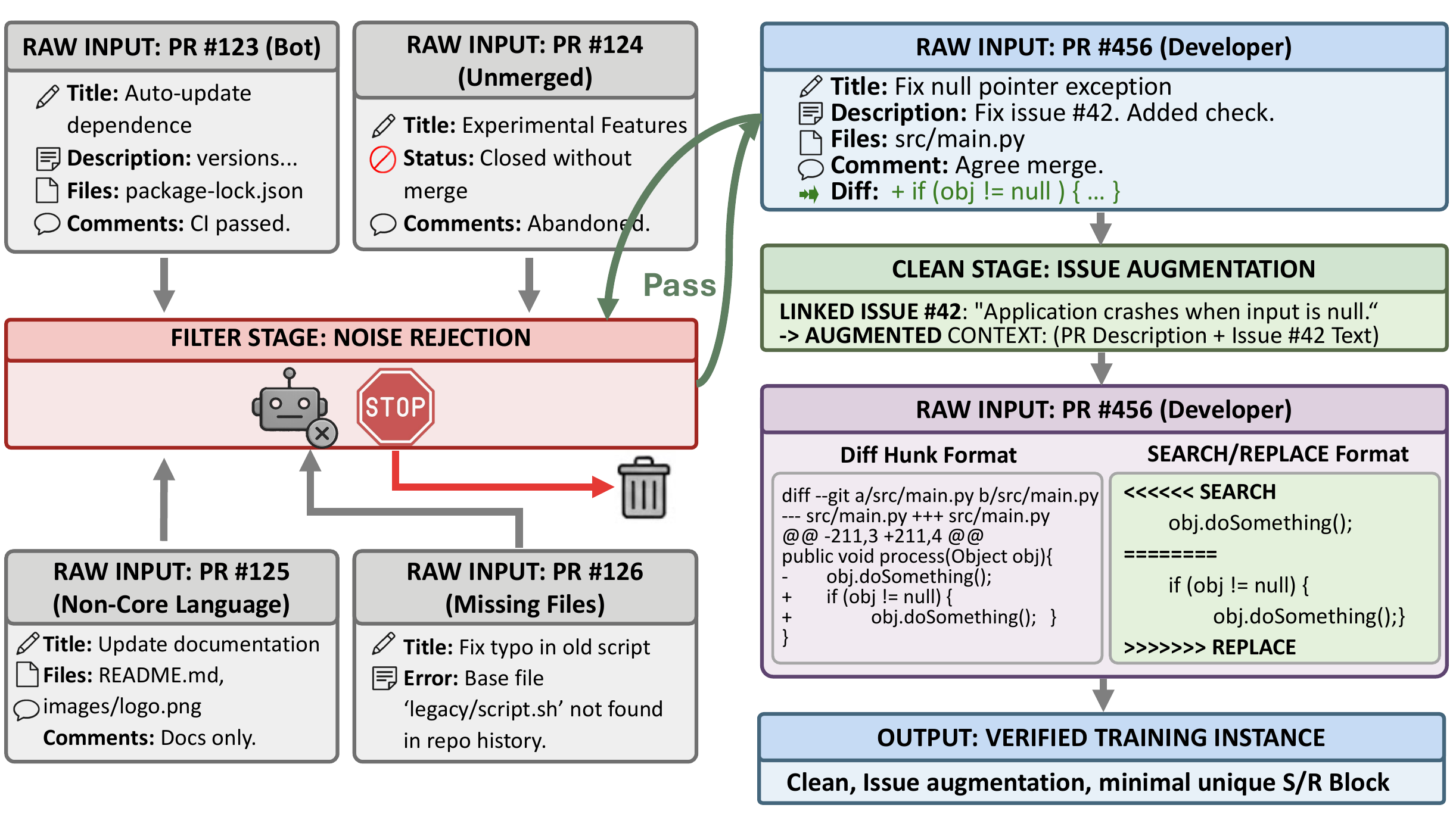}
\caption{\textbf{The Life of a Data Point: From Raw Noise to Verified Signal.} 
\textbf{Track A (Left)} illustrates the aggressive pruning of noise, rejecting inputs due to bot activity, unmerged status, non-core language files, or missing history. 
\textbf{Track B (Right)} depicts the transformation of a valid PR: it is \textbf{augmented} with the linked Issue context to recover user intent and \textbf{converted} into a deterministic \textit{Search/Replace} block for verifiable training.}
\label{fig:appendix_pipeline_visual}
\end{figure*}

\section{Data Processing Details}
\label{app:data_processing}

In this appendix, we provide the comprehensive implementation details of the data cleaning, filtering, reconstruction, and verification pipeline described in Section~\ref{sec:data}. The pipeline is implemented to ensure that only high-quality, reproducible, and semantic code changes are included in the \model dataset. Figure~\ref{fig:appendix_pipeline_visual} illustrates how \model\ bridges the ``noise-validity gap'' by strictly filtering failure modes and enhancing valid signals.

\subsection{Language Detection and Extension Rules}
\label{app:language_rules}

To ensure the quality of training data, we enforce a strict two-stage filtering pipeline based on file extensions.
First, we determine the primary programming language of each Pull Request (PR) by counting the modified files that match the \textbf{Core} extensions (e.g., \texttt{.py} for Python, \texttt{.java} for Java). The language with the highest frequency of core files is assigned to the PR. If a PR modifies no core files (e.g., only documentation changes), it is immediately discarded.

Second, to eliminate noise from binary files or unrelated assets, we apply a rigorous purity check using the \textbf{Allowed} set. Once the language is determined, we verify that \emph{every} file modified in the PR possesses an extension listed in the \emph{Allowed} set for that language (enumerated in Table~\ref{tab:language_rules}).
Crucially, this is a PR-level constraint: if a PR contains any file not listed in the Allowed set, the entire PR is dropped.
This ensures that our model is not exposed to PRs containing ambiguous or non-textual artefacts.
Finally, for valid PRs, we retain only the files matching the \emph{Core} extensions for training to focus on code logic changes.

\subsection{PR Validity and Noise Filtering}
\label{app:noise_filtering}

To distil high-quality editing signals from noisy GitHub data, we apply a multi-stage filtering pipeline. A PR is discarded if it triggers any of the following exclusion criteria:

\paragraph{1. Automation and Bot Filtering.}
We exclude PRs created by automated tools or where the only activity comes from bots. A user is classified as a bot if their username matches any of the following regular expressions:
\begin{itemize}
    \item \textbf{Suffix Patterns:} \texttt{bot\$}, \texttt{\_bot\$}, \texttt{-bot\$}
    \item \textbf{Prefix Patterns:} \texttt{\^{}bot}
    \item \textbf{Specific Services:} \texttt{dependabot}, \texttt{renovate}, \texttt{github-actions}, \texttt{travis-ci}, \texttt{circleci}, \texttt{coveralls}, \texttt{auto}, \texttt{automated}
\end{itemize}

\paragraph{2. Metadata and Quality Constraints.}
We filter PRs based on status and textual content to ensure semantic relevance:
\begin{itemize}
    \item \textbf{Status:} The PR must be marked as \texttt{MERGED} or \texttt{APPROVED}.
    \item \textbf{Title Blocklist:} We remove maintenance PRs containing keywords: \texttt{bump}, \texttt{dependencies}, \texttt{dependency}, \texttt{depend}, \texttt{release}.
    \item \textbf{Description Blocklist:} We remove descriptions containing \texttt{qwiet} (indicating automated security scans).
    \item \textbf{Length Heuristics:} To ensure sufficient context, Titles must be $\ge 10$ characters, and Descriptions must be $\ge 20$ characters.
\end{itemize}

\paragraph{3. Structural Integrity Checks.}
We strictly enforce that the PR represents a clean, in-place modification of existing code. We exclude PRs that:
\begin{itemize}
    \item \textbf{Missing Base Code or Diff:} PRs with empty base code files or missing diffs are filtered out.
    
    \item \textbf{Mismatched Files:} Have a discrepancy between the set of files in the base state and the files in the diff (i.e., bijective mapping required).
\end{itemize}

\subsection{Bug Identification and Issue Linking}
\label{app:bug_linking}

We augment PRs with context from linked Issues. We extract issue numbers from titles and descriptions using the following prioritised regex patterns:

\begin{enumerate}[noitemsep]
    \item \textbf{Hash References:} \texttt{\#(\textbackslash d+)}
    \item \textbf{Explicit Keywords:}
    \begin{itemize}
        \item \texttt{issue[:\textbackslash s\#-]*(\textbackslash d+)}
        \item \texttt{bug[:\textbackslash s\#-]*(\textbackslash d+)}
        \item \texttt{fix(es)?[:\textbackslash s\#-]*(\textbackslash d+)}
        \item \texttt{resolve(s|d)?[:\textbackslash s\#-]*(\textbackslash d+)}
        \item \texttt{close(s|d)?[:\textbackslash s\#-]*(\textbackslash d+)}
        \item \texttt{gh-(\textbackslash d+)}
    \end{itemize}
\end{enumerate}

\subsection{Verified Search/Replace Conversion Pipeline}
\label{app:sr_conversion}

\begin{algorithm}[h]
    \SetKwInOut{Input}{Input}\SetKwInOut{Output}{Output}
    \Input{Base File Content $C_{base}$, Raw Diff Hunk $D$}
    \Output{Set of Verified Blocks $\mathcal{S}$ or Failure $\bot$}

    \tcp{\small Phase 1: Ground Truth Reconstruction}
    $C_{target} \gets \text{FakeGitApply}(C_{base}, D)$\;
    
    \If{$C_{target}$ is Invalid}{
        \textbf{return} $\bot$
    }
    \tcp{\small Phase 2: Minimal Unique Context Search}
    $\Delta \gets \text{ComputeDiffOps}(C_{base}, C_{target})$\;
    
    $\mathcal{S} \gets \emptyset$\;

    \For{edit operation $\delta \in \Delta$}{
        Let $[s, e]$ be the line range of $\delta$ in $C_{base}$\;
        
        $k \gets 0$ \tcp{\small Init context size}
        \For{$k$ \textbf{in} range(0, MAX\_CONTEXT)}{
            \tcp{\small Expand window symmetrically}
            $start \gets \max(0, s - \lfloor k/2 \rfloor)$\;
            
            $end \gets \min(\text{Len}(C_{base}), e + \lceil k/2 \rceil)$\;
            
            $S_{search} \gets C_{base}[start : end]$\;
            
            \tcp{Check uniqueness in full file}
            \If{$C_{base}.\text{Count}(S_{search}) == 1$}{
                $S_{replace} \gets \text{GetNewContent}(C_{target}, \delta)$\;
                
                $\mathcal{S}.\text{add}(\{S_{search}, S_{replace}\})$\;
                
                \textbf{break}
            }
        }
        
    }

    \tcp{\small Phase 3: Round-Trip Verification}
    $C_{verify} \gets C_{base}$\;
    
    \For{each block $B \in \mathcal{S}$}{
        \tcp{Deterministic String Replacement}
        $locs \gets \text{FindIndices}(B.search, C_{verify})$\;

        \If{$\text{Length}(locs) \neq 1$}{
            \textbf{return} $\bot$ \tcp{Safety check failed}
        }
        $C_{verify} \gets \text{Replace}(C_{verify}, B.search, B.replace)$\;
    }

    \tcp{Bit-wise Equality Check}\;
    \If{$C_{verify} == C_{target}$}{
        \textbf{return} $\mathcal{S}$\;
    }
    \Else{
        \textbf{return} $\bot$ \tcp{Artefacts detected}
    }    
\caption{Specification Compatibility Checking}\label{alg:sr_gen_detailed}

\end{algorithm}

We convert raw unified diffs into deterministically verifiable \textbf{Search/Replace} blocks through a three-stage pipeline, as detailed in Algorithm~\ref{alg:sr_gen_detailed}.

\paragraph{1. Ground-Truth Reconstruction (Fake Git Apply).}
We reconstruct the ``After'' state using a git sandbox to handle context fuzzy-matching.
\begin{enumerate}[noitemsep]
    \item Initialise a temporary git repository with the base code.
    \item Apply the diff hunk using \texttt{git apply} with fallback strategies:
    \begin{itemize}
        \item \texttt{--verbose}
        \item \texttt{--ignore-whitespace}
        \item \texttt{--ignore-space-change}
        \item \texttt{--whitespace=fix}
    \end{itemize}
    \item If the application fails, the PR is discarded. If successful, the result defines the \texttt{expected\_content}.
\end{enumerate}

\paragraph{2. Minimal Unique Context Search.}
We identify edit spans by computing the difference between the base and the reconstructed target files. To generate \texttt{SEARCH} blocks that are both concise and unambiguous, we employ an iterative expansion strategy:
\begin{itemize}
    \item \textbf{Edit Merging:} Adjacent edits (separated by $\le 1$ line) are coalesced into a single block to maintain semantic continuity.
    \item \textbf{Context Expansion:} For each edit, we initialise a context window of size zero. We iteratively expand this window symmetrically: adding lines above and below the edit until the resulting \texttt{SEARCH} block occurs \textbf{exactly once} within the full file. This guarantees that the model learns the \textit{minimum} context necessary for unique localisation.
\end{itemize}

\paragraph{3. Round-Trip Verification.}
We validate the generated blocks by performing a strict ``round-trip'' application using simple string replacement, independent of git. A training instance is retained only if it passes three integrity checks:
\begin{enumerate}[noitemsep]
    \item \textbf{Uniqueness:} Each generated \texttt{SEARCH} block must be found exactly once in the base file.
    \item \textbf{Non-Overlapping:} Multiple edit blocks within the same file must not have overlapping search regions.
    \item \textbf{Exact Reconstruction:} Applying the Search/Replace blocks to the base file via string replacement must yield a file that is \textbf{bit-wise identical} to the ground-truth \texttt{target\_content} derived in Step 1.
\end{enumerate}

\subsection{Context Windowing Strategy}
\label{app:windowing}

For files exceeding the token limit (e.g., 100k tokens), we employ a focus-and-expand strategy:
\begin{enumerate}[noitemsep]
    \item \textbf{Identify Ranges:} Extract line ranges $[start, end]$ covered by verified Search/Replace blocks.
    \item \textbf{Expand:} Extend each range by $N=20$ lines to capture local definitions.
    \item \textbf{Merge \& Reconstruct:} Merge overlapping ranges and concatenate them, inserting  markers for omitted sections.
\end{enumerate}
This ensures the model sees the necessary context for the edit without processing the entire file.

\subsection{Rigorous Decontamination Protocol}
\label{app:decontamination_detailed}

To ensure the integrity of our evaluation on SWE-bench and address potential leakage via code propagation (e.g., forks, vendored dependencies), we enforce a multi-layered decontamination pipeline.

\paragraph{1. Repository-Level Exclusion.}
As a primary defence, we strictly blocklist all repositories present in the SWE-bench Lite and Verified metadata. Any Pull Request originating from or targeting these repositories is structurally discarded.

\paragraph{2. Content-Based Decontamination (Addressing Code Movement).}
Relying solely on repository names is insufficient due to the prevalence of code cloning and vendored directories. To mitigate this, we implement content-aware filtering:
\begin{itemize}
    \item \textbf{Exact File Matching:} We compute SHA-256 hashes for all source files in the training corpus. If any file strictly matches a file version found in the evaluation set (spanning the entire test timeline), the instance is flagged. This effectively catches copied or moved code regardless of the repository it resides in.
    \item \textbf{N-gram Overlap:} For partial matches, we index all \textit{Gold Patches} and \textit{Issue Descriptions} from the test set. We exclude training instances that share a \textbf{15-gram} code subsequence with gold patches or exceed a \textbf{0.5 Jaccard similarity} with issue descriptions, following established protocols~\citep{ kocetkov2022stack}.
\end{itemize}

\subsection{Semantic Leakage Analysis}
\label{app:semantic_leakage}

Beyond lexical decontamination, we further test whether residual train-test similarity explains \model's SWE-bench Verified performance. Following the contamination analysis framework of \citet{riddell2024quantifying}, we evaluate the hypothesis that leaked or near-leaked examples should make more similar test instances easier to solve.

For each Clean-PR training sample, we concatenate the issue description and code patch; for each SWE-bench Verified instance, we concatenate the problem statement and gold patch. We embed the Python subset of \model-train (approximately 390K samples) and all 500 SWE-bench Verified instances using BGE-Code-v1~\citep{li2025coder}, build a FAISS IndexFlatIP index~\citep{johnson2021billion}, and assign each test instance its maximum nearest-neighbour similarity score. We then split the evaluation set into five equal-sized similarity quintiles.

\begin{table}[h]
\caption{SWE-bench Verified resolution rate by maximum train-test similarity quintile.}
\label{tab:semantic_leakage}
\centering
\resizebox{\linewidth}{!}{
\begin{tabular}{lccc}
\toprule
\textbf{Similarity Quintile} & \textbf{Sim Range} & \textbf{\# Inst.} & \textbf{Resolve Rate} \\
\midrule
Q5 (most similar) & [0.76, 0.94) & 100 & 25.0 \\
Q4 & [0.72, 0.76) & 100 & 33.0 \\
Q3 & [0.67, 0.72) & 100 & 34.0 \\
Q2 & [0.61, 0.67) & 100 & 30.0 \\
Q1 (least similar) & [0.35, 0.61) & 100 & 31.0 \\
\midrule
\textbf{Overall} & -- & \textbf{500} & \textbf{30.6} \\
\bottomrule
\end{tabular}
}
\end{table}

The pattern does not support the leakage hypothesis: the most similar quintile has the lowest resolve rate (25.0\%), while the least similar quintile reaches 31.0\%. The Pearson correlation between maximum similarity and binary resolve outcome is $r=-0.061$ with $p=0.184$, which is not statistically significant. We also manually inspected the top-10 most similar train-test pairs and found common coding idioms or library-level patterns rather than task-specific leakage. Combined with the repository exclusion, SHA-256 matching, 15-gram code filtering, and issue Jaccard filtering above, this suggests that the observed gains are better explained by generalisation than by contamination.

\begin{table}[h]
\caption{Language distribution for \textbf{\model-full} (Pre-sampling).}
\label{tab:lang_dist_full}
\centering
\resizebox{1\linewidth}{!}{
\begin{tabular}{lrrr}
\toprule
\textbf{Language} & \textbf{Count} & \textbf{Ratio (\%)} & \textbf{Tokens (B)} \\
\midrule
Python & 543,419 & 17.81 & 7.77 \\
C++ & 235,246 & 7.71 & 7.45 \\
Go & 409,859 & 13.43 & 6.80 \\
Java & 454,981 & 14.91 & 6.17 \\
JavaScript & 371,640 & 12.18 & 4.55 \\
Rust & 239,346 & 7.85 & 4.12 \\
TypeScript & 278,881 & 9.14 & 3.07 \\
C & 81,789 & 2.68 & 2.29 \\
Kotlin & 132,316 & 4.34 & 1.15 \\
C\# & 88,990 & 2.92 & 1.11 \\
PHP & 64,526 & 2.12 & 0.96 \\
Ruby & 149,946 & 4.91 & 0.94 \\
\midrule
\textbf{Total} & \textbf{3,050,939} & \textbf{100.00} & \textbf{46.38} \\
\bottomrule
\end{tabular}
}
\end{table}

\begin{table}[h]
\caption{Language distribution for \textbf{\model-train} (Post-sampling). This dataset is used for mid-training.}
\label{tab:lang_dist_train}
\centering
\resizebox{1\linewidth}{!}{
\begin{tabular}{lrrr}
\toprule
\textbf{Language} & \textbf{Count} & \textbf{Ratio (\%)} & \textbf{Tokens (B)} \\
\midrule
Python & 389,881 & 19.34 & 3.83 \\
Go & 268,302 & 13.31 & 2.33 \\
C++ & 154,346 & 7.66 & 2.33 \\
JavaScript & 269,176 & 13.35 & 2.04 \\
Java & 248,251 & 12.32 & 1.91 \\
Rust & 150,024 & 7.44 & 1.52 \\
TypeScript & 188,690 & 9.36 & 1.22 \\
C & 56,812 & 2.82 & 0.76 \\
Ruby & 109,640 & 5.44 & 0.54 \\
C\# & 58,045 & 2.88 & 0.45 \\
Kotlin & 78,238 & 3.88 & 0.40 \\
PHP & 44,303 & 2.20 & 0.35 \\
\midrule
\textbf{Total} & \textbf{2,015,708} & \textbf{100.00} & \textbf{17.67} \\
\bottomrule
\end{tabular}
}
\end{table}

\subsection{Language Distribution}
\label{app:Language_Distribution}
We support 12 major programming languages. Table~\ref{tab:lang_dist_full} and  Table~\ref{tab:lang_dist_train} detail the distribution of instances and tokens for the Full and Train sets, respectively. The filtering process preserves the relative diversity of languages, with Python, Go, and C++ remaining the dominant contributors.

\subsection{Data Formatting}

\paragraph{Input Sequence Template.}
Table~\ref{tab:input_template} illustrates the exact string formatting template used to construct the Mid-training sequences. We linearise the repository context, issue description, and code base into a unified text stream, followed by the target Search/Replace edits.

\begin{table}[!htbp]
    \centering
    \caption{The linearised input template used for Mid-training.}
    \label{tab:input_template}
    \begin{tcolorbox}[ width=\linewidth, arc=2mm, title={\textbf{\model Format}}]
\begin{verbatim}
Repository Name: {repo_name}
Pull Request title: {pr_title}
Description: 
{pr_description}

Pull Request codes:
{base_code_content}

SEARCH/REPLACE edits:
{search_replace_format}

Comments:
{valid_comments}
\end{verbatim}
    \end{tcolorbox}
\end{table}

\subsection{Data Release Specifications}
\label{app:data_release_specs}

To ensure full reproducibility and facilitate downstream analysis, we will release the \model dataset with comprehensive metadata. 
Table~\ref{tab:data_schema} details the definition of each field, including repository metadata, statistical metrics (e.g., token counts), and processing flags (e.g., windowing usage).

\begin{table*}[!htbp]
    \centering
    \small
    \caption{Detailed schema of the released \model dataset. The corpus retains granular metadata and statistics to support diverse research directions beyond direct training.}
    \label{tab:data_schema}
    \renewcommand{\arraystretch}{1.2}
    \resizebox{1\textwidth}{!}{%
    \begin{tabularx}{1 \textwidth}{l|l|X}
        \toprule
        \textbf{Category} & \textbf{Field Name} & \textbf{Description} \\
        \midrule
        \multirow{4}{*}{\textbf{Metadata}} 
         & \texttt{repo\_name} & The identifier of the source repository (e.g., \texttt{owner/repo}). \\
         & \texttt{repo\_url} & The persistent URL to the GitHub repository for attribution. \\
         & \texttt{detected\_language} & The primary programming language of the modified files (e.g., Python). \\
         & \texttt{is\_use\_windows} & Boolean flag indicating if the base code was truncated/windowed. \\
        \midrule
        \multirow{3}{*}{\textbf{Content}} 
         & \texttt{pr\_title} & The original title of the Pull Request. \\
         & \texttt{pr\_description} & The detailed issue description or PR body text outlining the intent. \\
         & \texttt{formatted\_text} & The final flattened string sequence constructed using the template in Table~\ref{tab:input_template}. \\
        \midrule
        \multirow{3}{*}{\textbf{Code Artefacts}}
         & \texttt{base\_code} & The raw content of the source files \textit{before} the edits are applied. \\
         & \texttt{diff} & The verified \textbf{Search/Replace} block sequence used as the training target. \\
         & \texttt{valid\_comments} & (Optional) Reviewer comments aligned with the code changes, if available. \\
        \midrule
        \multirow{3}{*}{\textbf{Statistics}} 
         & \texttt{token\_count} & The total number of tokens in the \texttt{formatted\_text} (using Qwen2.5 Coder tokenizer). \\
         & \texttt{changed\_files\_count} & The number of distinct files modified in this Pull Request. \\
         & \texttt{diff\_lines} & The total number of lines added or removed in the diff hunk. \\
        \bottomrule
    \end{tabularx}%
    }
\end{table*}

\section{StarCoder2-style Data Construction}
\label{app:starcoder2_baseline}

To ensure a fair comparison, we constructed a strong baseline dataset rigorously following the data processing pipeline of StarCoder2~\citep{lozhkov2024starcoder2stackv2}. Starting from our raw collection of \textbf{16.4 million} crawled Pull Requests (PRs), we applied a multi-stage filtering, sampling, and formatting protocol.

\subsection{Filtering and Cleaning Pipeline}
We implemented a cascade of filters targeting PR metadata, file content, and text quality.

\paragraph{PR-level Filtering.}
We discard PRs that satisfy any of the following criteria:
\begin{itemize}
    \item \textbf{Bot Activity:} PRs opened by bots or containing comments exclusively from bots (identified by username patterns and keywords).
    \item \textbf{Licence \& Status:} PRs from repositories with non-permissive licences (e.g., GPL), user opt-outs, or PRs that were not approved or merged.
    \item \textbf{Integrity:} PRs that change the base branch during the process or lack initial diffs, preventing accurate reconstruction of changes.
\end{itemize}

\paragraph{File-level Filtering.}
For the files involved in each PR, we apply strict quality controls:
\begin{itemize}
    \item \textbf{Size Constraints:} Files exceeding 1MB in size, 100,000 lines, an average line length $>100$, or a maximum line length $>1,000$ are removed.
    \item \textbf{Content Quality:} Files with $<25\%$ alphanumeric characters or $>25\%$ hexadecimal characters are discarded to remove binary or obfuscated files. Non-English Markdown files are also excluded.
\end{itemize}

\paragraph{Text Cleaning.}
To ensure high-quality natural language supervision:
\begin{itemize}
    \item \textbf{Length \& Keywords:} We remove PRs with titles $<10$ characters (or containing generic terms like ``dependency'', ``release'') and descriptions $<20$ characters (or containing spam keywords like ``Qwiet'').
    \item \textbf{Truncation:} Titles are truncated to 500 characters. Descriptions are truncated to 80 lines (preserving the first 60 and last 20 lines) or a maximum of 1,000 characters.
    \item \textbf{Comment Sanitization:} We remove auto-generated email replies. Comments shorter than 20 characters are discarded unless they are code review comments. For review comments, associated diff hunks $>10,000$ characters are truncated. All usernames are anonymized to identifiers like \texttt{username\_0}.
\end{itemize}

\textbf{Result:} After this rigorous filtering, the dataset was reduced from 16.4M to \textbf{6,037,781} valid PRs (a 36.8\% pass rate).

\paragraph{Pull Request Template.}
The PR input sequence is constructed as follows:
\begin{tcolorbox}[colback=gray!5,  title=\textbf{StarCoder2-style PR Format}, fontupper=\small\ttfamily]
Pull Request Title: \{title\}

Created by username\_0: \{description\}
Status: \{status\}

Repository Name: \{repo\_name\}

Base files:

File: \{filepath\}

Content:
\{content\}

Diff changes: \{diffchange\}

Comments:

Comment by \{username\}: \{content\}
\end{tcolorbox}

\subsection{Rebalancing and Sampling}
To mitigate the over-representation of prolific repositories, we adopt the linear downsampling strategy used in StarCoder2. Concretely, for a repository containing $n$ valid PRs, we retain PRs with a probability that depends on $n$: when $n=1$, the retention probability is set to $0.8$; when $1 < n \le 1000$, the probability decreases linearly from $0.8$ to $0.1$ as $n$ increases; and when $n>1000$, we set the probability so that, in expectation, exactly 100 PRs are retained from that repository. After applying this sampling procedure, the dataset is further reduced to \textbf{2,112,688} high-quality PR instances.

\subsection{Data Formatting}
We serialise the PRs and Issues into a unified text format. Unlike our proposed method which uses explicit Search/Replace blocks, the StarCoder2-style baseline uses a descriptive natural language format.

\paragraph{Issue Template.}
We also aggregate linked GitHub Issues using the standard conversation format:
\begin{tcolorbox}[colback=gray!5, title=\textbf{StarCoder2-style Issue Format}, fontupper=\small\ttfamily]
Title: \{title\}

Issue: \{issue\_content\}
\end{tcolorbox}

The final StarCoder2-style baseline dataset comprises \textbf{17.4 billion} tokens.

\begin{table}[h]
\caption{Training configurations for mid-training and SFT.}
\label{tab:train_config}
\centering
\resizebox{1\linewidth}{!}{
\begin{tabular}{lcc}
\toprule
\textbf{Setting} & \textbf{Mid-training} & \textbf{SFT} \\
\midrule
Model size & 32B & 32B \\
Precision & BF16 & BF16 \\
DeepSpeed ZeRO-3 & \cmark & \cmark \\
FlashAttention-2 & \cmark & \cmark \\
Liger-Kernel & \cmark & \cmark \\
Optimiser & AdamW & AdamW \\
LR scheduler & Cosine & Cosine \\
Warmup ratio & 0.03 & 0.03 \\
Peak learning rate & $2.0 \times 10^{-5}$ & $5.0 \times 10^{-6}$ \\
Epochs & 2 & 3 \\
Global batch size & 128 & 128 \\
Per-device batch size & 2 & 2 \\
Gradient accumulation & 2 & 2 \\
GPU type & H200  & H200 \\
GPU counts & 32& 32 \\
Context length & 32{,}768 & 32{,}768 \\
Training time (wall-clock) & 259 h & 38 h \\
\bottomrule
\end{tabular}
}
\end{table}

\section{Training and inference configuration}
\label{sec:appendix_prompts}

\paragraph{Inference Framework.}
We adopt a \textbf{Simplified Agentless} scaffolding for evaluation, which mirrors our training alignment by decomposing the resolution process into three deterministic steps: (1) \textbf{File localisation}~(Table~\ref{tab:prompt_step1}), (2) \textbf{Line-level navigation}~(Table~\ref{tab:prompt_step2}), and (3) \textbf{Patch Generation}~(Table~\ref{tab:prompt_step3}).
We utilise default decoding parameters (greedy decoding with temperature $0$) to ensure reproducibility. For the experiments in Section~\ref{sec:pass@k_part}, we set the temperature to 0.8.
Crucially, to optimise the context window usage, we enforce strict retrieval constraints: for the downstream \textbf{Context Construction} (Step 2) and \textbf{Patch Generation} (Step 3) phases, we only retain the \textbf{top-3} ranked files identified in the initial localisation step.

\paragraph{Training configurations.}
\label{app:training_details}
We train the 32B model with BF16 using DeepSpeed ZeRO-3, FlashAttention-2, and Liger-Kernel optimisations (Table~\ref{tab:train_config}).
For mid-training, we use AdamW with a cosine learning-rate schedule and a warmup ratio of 0.03, training for 2 epochs with a global batch size of 128 (per-device batch size 2 with 2 gradient-accumulation steps) and a peak learning rate of $2.0 \times 10^{-5}$.
The SFT stage inherits the same hardware configuration and context length, but uses a smaller learning rate of $5.0 \times 10^{-6}$ for stable adaptation.

\section{Discussion and Future Work}
\label{sec:discussion}

Our at-most-five-core-files filter retains the majority of verified PRs but under-represents the long tail of broad, cross-cutting repository changes; handling such large edits will require longer-context training data and stronger planning or verification components. Extending \model to other architectures, tokenizers, and MoE-based frontier backbones is a natural next step but lies beyond the scope of this work.

\begin{table}[!htbp]
    \centering
    \small
    \caption{Prompt for Step 1: File localisation}
    \label{tab:prompt_step1}
    \begin{tcolorbox}[width=.49\textwidth, title={\textbf{Prompt for File localisation}}]
        Please look through the following GitHub problem description and Repository structure and provide a list of files that one would need to edit to fix the problem.\\
        \\
        \textbf{\#\#\# GitHub Problem Description \#\#\#}\\
        \textbf{\{problem\_statement\}}\\
        \\
        \textbf{\#\#\#}\\
        \\
        \textbf{\#\#\# Repository Structure \#\#\#}\\
        \textbf{\{structure\}}\\
        \\
        \textbf{\#\#\#}\\
        \\
        Please only provide the full path and return at most 5 files.\\
        The returned files should be separated by new lines ordered by most to least important and wrapped with \texttt{```}\\
        For example:\\
        \texttt{```}\\
        file1.py\\
        file2.py\\
        \texttt{```}
    \end{tcolorbox}
\end{table}

\begin{table*}[!htbp]
    \centering
    \small
    \caption{Prompt for Step 2: Fine-grained Navigation}
    \label{tab:prompt_step2}
    \begin{tcolorbox}[width=\textwidth, title={\textbf{Prompt for Fine-grained Navigation}}]
        Please review the following GitHub problem description and relevant files, and provide a set of locations that need to be edited to fix the issue.\\
        The locations can be specified as class names, function or method names, or exact line numbers that require modification.\\
        \\
        \textbf{\#\#\# GitHub Problem Description \#\#\#}\\
        \textbf{\{problem\_statement\}}\\
        \\
        \textbf{\#\#\#}\\
        \textbf{\{file\_contents\}}\\
        \\
        \textbf{\#\#\#}\\
        \\
        Please provide the class name, function or method name, or the exact line numbers that need to be edited.\\
        The possible location outputs should be either "class", "function" or "line".\\
        \\
        \textbf{\#\#\# Examples:}\\
        \texttt{```}\\
        full\_path1/file1.py\\
        line: 10\\
        class: MyClass1\\
        line: 51\\
        \\
        full\_path2/file2.py\\
        function: MyClass2.my\_method\\
        line: 12\\
        \\
        full\_path3/file3.py\\
        function: my\_function\\
        line: 24\\
        line: 156\\
        \texttt{```}\\
        \\
        Return just the location(s) wrapped with \texttt{```}.
    \end{tcolorbox}
\end{table*}

\begin{table*}[!htbp]
    \centering
    \small
    \caption{Prompt for Step 3: Patch Generation}
    \label{tab:prompt_step3}
    \begin{tcolorbox}[width=\textwidth, title={\textbf{Prompt for Patch Generation}}]
        We are currently solving the following issue within our repository. Here is the issue text:\\
        --- BEGIN ISSUE ---\\
        \textbf{\{problem\_statement\}}\\
        --- END ISSUE ---\\
        \\
        \textbf{\{repair\_relevant\_file\_instruction\}}\\
        --- BEGIN FILE ---\\
        \texttt{```}\\
        \textbf{\{content\}}\\
        \texttt{```}\\
        --- END FILE ---\\
        \\
        Please first localise the bug based on the issue statement, and then generate *SEARCH/REPLACE* edits to fix the issue.\\
        \\
        Every *SEARCH/REPLACE* edit must use this format:\\
        1. The file path\\
        2. The start of search block: \texttt{<<<<<<< SEARCH}\\
        3. A contiguous chunk of lines to search for in the existing source code\\
        4. The dividing line: \texttt{=======}\\
        5. The lines to replace into the source code\\
        6. The end of the replace block: \texttt{>>>>>>> REPLACE}\\
        \\
        Here is an example:\\
        \\
        \texttt{```python}\\
        \#\#\# mathweb/flask/app.py\\
        \texttt{<<<<<<< SEARCH}\\
        from flask import Flask\\
        \texttt{=======}\\
        import math\\
        from flask import Flask\\
        \texttt{>>>>>>> REPLACE}\\
        \texttt{```}\\
        \\
        Please note that the *SEARCH/REPLACE* edit REQUIRES PROPER INDENTATION. If you would like to add the line '        print(x)', you must fully write that out, with all those spaces before the code!\\
        Wrap the *SEARCH/REPLACE* edit in blocks \texttt{```python...```}.
    \end{tcolorbox}
\end{table*}

\end{document}